\documentclass{SciPost}

\hypersetup{
    colorlinks,
    linkcolor={red!50!black},
    citecolor={blue!50!black},
    urlcolor={blue!80!black}
}

\usepackage[bitstream-charter]{mathdesign}
\DeclareSymbolFont{usualmathcal}{OMS}{cmsy}{m}{n}
\DeclareSymbolFontAlphabet{\mathcal}{usualmathcal}

\fancypagestyle{SPstyle}{
\fancyhf{}
\lhead{\colorbox{scipostblue}{\bf \color{white} ~SciPost Physics }}
\rhead{{\bf \color{scipostdeepblue} ~Submission }}

\fancyfoot[C]{\textbf{\thepage}}
}

\usepackage[normalem]{ulem}

\begin{document}

\pagestyle{SPstyle}

\begin{center}{\Large \textbf{\color{scipostdeepblue}{
On the upper critical dimension of the KPZ universality class:\\ KPZ and related equations on a fully connected graph
\\
}}}\end{center}

\begin{center}\textbf{
J. M. Marcos\textsuperscript{1,2$\star$},
J. J. Mel\'endez\textsuperscript{1,2},
R. Cuerno\textsuperscript{3} and
J. J. Ruiz-Lorenzo\textsuperscript{1,2}
}\end{center}

\begin{center}
{\bf 1} Departamento de F\'{\i}sica, Universidad de Extremadura, 06006 Badajoz, Spain
\\
{\bf 2} Instituto de Computaci\'on Cient\'{\i}fica Avanzada de Extremadura (ICCAEx), Universidad de Extremadura, 06006 Badajoz, Spain
\\
{\bf 3} Universidad Carlos III de Madrid, Departamento de Matem\'aticas and Grupo Interdisciplinar de Sistemas Complejos (GISC), Avenida de la Universidad 30, 28911 Legan\'es (Madrid), Spain
\\[\baselineskip]
$\star$ \href{mailto:email1}{\small jesusmm@unex.es}
\end{center}

\section*{\color{scipostdeepblue}{Abstract}}
\textbf{\boldmath{%
We investigate the infinite-dimensional limit of nonequilibrium surface growth by numerically integrating stochastic growth equations on a fully connected graph. In particular, we study the Edwards–Wilkinson (EW), Kardar–Parisi–Zhang (KPZ), and tensionless KPZ (TKPZ) equations. Using a network discretization adapted to the all-to-all interaction topology, we analyze the global roughness, height-fluctuation statistics, time power spectra, and two-time correlations. For the EW equation, we obtain an exact expression for the roughness that matches the numerical simulations and shows that the interface becomes flat as the number of nodes in the graph, $N$, tends to infinity. We also compute analytically the time power spectrum, show that height fluctuations are Gaussian, and derive an explicit expression for the two-time height autocorrelation function, indicating that the aging behavior is trivial. For the KPZ equation, finite-size and strong-coupling effects can cause deviations from EW behavior at moderate system sizes $N$, often accompanied by numerical instabilities; however, these differences disappear as $N$ increases. In the large-$N$ limit, KPZ dynamics converges to EW behavior, as the four observables analyzed exhibit identical scaling properties. Overall, our results indicate that on a fully connected graph the KPZ nonlinearity is irrelevant as $N\to\infty$, leading to EW-like dynamics with asymptotically flat interfaces. These findings are interpreted in the context of the upper critical dimension of the KPZ universality class.
}}

\vspace{\baselineskip}



\vspace{10pt}
\noindent\rule{\textwidth}{1pt}
\tableofcontents
\noindent\rule{\textwidth}{1pt}
\vspace{10pt}


\section{Introduction}\label{sec:intro}

The Kardar--Parisi--Zhang (KPZ) equation \cite{Kardar1986} provides a fundamental framework for describing the large-scale, scale-invariant dynamics observed in a broad class of non-equilibrium growth phenomena \cite{Barabasi1995,Krug1997,Takeuchi2018}. It governs the temporal evolution of a fluctuating interface and is given by
\begin{equation}
    \frac{\partial h}{\partial t}(\boldsymbol{x},t)
    = \nu \nabla^2 h(\boldsymbol{x},t)
    + \frac{\lambda}{2} \left[ \nabla h(\boldsymbol{x},t) \right]^2
    + \eta(\boldsymbol{x},t)\, .
    \label{eq:KPZ_orig}
\end{equation}
In this expression, $h(\boldsymbol{x},t)$ denotes the interface height measured from a reference substrate at position $\boldsymbol{x} \in \mathbb{R}^d$ and time $t$. The Laplacian term, weighted by the coefficient $\nu$, represents relaxation mechanisms associated with surface tension, while the nonlinear gradient term, with strength $\lambda$, accounts for growth occurring along the local normal to the interface. The stochastic contribution $\eta(\boldsymbol{x},t)$ introduces random fluctuations and is modeled as Gaussian white noise with zero mean, $\langle \eta(\boldsymbol{x},t) \rangle = 0$, and amplitude $2D$ so that
\begin{equation}
\label{eq:ruido}
\langle \eta(\boldsymbol{x},t)\,\eta(\boldsymbol{x}',t') \rangle
= 2D\,\delta^d(\boldsymbol{x} - \boldsymbol{x}')\,\delta(t - t')\,,
\end{equation}
where $\delta^d(\cdot)$ denotes the Dirac delta function in dimension $d$. 

While the KPZ equation was originally introduced to describe interface growth, its range of applicability has expanded considerably in recent years. It is now recognized as a useful description in multiple areas of nonequilibrium physics, spanning from active matter systems \cite{Chen2016,Caballero2020} to synchronization \cite{Gutierrez2024}, or to quantum many-body dynamics \cite{Sieberer2025}. In addition, this stochastic evolution equation is closely connected to other continuum theories belonging to distinct universality classes. In particular, when the nonlinear coupling is set to zero, $\lambda = 0$, the model reduces to the Edwards--Wilkinson (EW) equation \cite{Edwards1982,Barabasi1995}, which corresponds to the Gaussian approximation of the stochastic time-dependent Ginzburg--Landau equation at criticality \cite{Kardar2007}. Conversely, eliminating surface tension by taking $\nu = 0$ leads to the so-called tensionless KPZ (TKPZ) equation \cite{Cartes2022,RodriguezFernandez2022,Fontaine2023}, whose relevance has been suggested for, among others, front dynamics in invasion percolation processes \cite{Asikainen2002,Asikainen2002b,RodriguezFernandez2022}, quantum matter \cite{Fujimoto2020,Fontaine2023,Vercesi2023}, or reaction-diffusion systems exhibiting phase turbulence \cite{Vercesi2024}.

The scale-invariant properties of these equations are encapsulated by the Family--Vicsek (FV) dynamic scaling ansatz for the surface width (or roughness) $w(L,t)$, see Eq.~\eqref{eq:width}, \cite{Barabasi1995,Krug1997}
\begin{equation}
	\label{eq:w}
	w(L,t)=t^{\beta} f\!\left(t/L^z\right),
\end{equation}
where $f(\cdot)$ is a scaling function such that the roughness grows as a power-law, $w \sim t^{\beta}$, for short times ($t \ll L^z$), and saturates as $w_{\mathrm{sat}} \sim L^{\alpha}$ for long times ($t \gg L^z$). Here, $\alpha$ is the roughness exponent, associated with the large-scale height fluctuations of the interface \cite{Barabasi1995,MozoLuis2022}, $\beta$ is the growth exponent, and $z$ is the dynamic exponent governing the growth of the lateral correlation length, $\xi(t) \sim t^{1/z}$. In the FV ansatz these exponents satisfy the relation $\alpha=\beta z$ \cite{Barabasi1995,Krug1997}.

For the EW equation, the scaling exponents can be obtained exactly in any spatial dimension $d$ \cite{Barabasi1995,Kardar2007}:
\begin{equation}
	\label{eq:exponentes_ew}
	\alpha=\frac{2-d}{2},\quad \beta=\frac{2-d}{4},\quad z=2.
\end{equation}
When $d < 2$, the interface obeys the FV scaling form. At the marginal dimension $d=2$, the (squared) roughness grows logarithmically in time prior to saturation, while its saturation value depends logarithmically on the system size. For dimensions $d>2$, the roughness exponent becomes negative and the interface remains asymptotically flat \cite{Barabasi1995}, meaning that the width no longer increases with system size $L$. Accordingly, the upper critical dimension of the EW universality class is $d^{\rm EW}_u=2$.

For the KPZ equation, exact analytical results for the scaling exponents are only available in one spatial dimension \cite{Barabasi1995,Krug1997,Takeuchi2018}, where
\begin{equation}
	\label{eq:exponentes_kpz_1d}
	\alpha=\frac{1}{2}, \quad \beta=\frac{1}{3}, \quad z=\frac{3}{2}.
\end{equation}
Moreover, the so-called Galilean scaling relation $\alpha + z = 2$ is expected to hold at the KPZ fixed point, in the renormalization group (RG) sense \cite{Kardar2007}, for all $d\leq d^{\rm KPZ}_u = d_u$, so that only one of the exponents remains independent. For $d>1$, exponents must be estimated through numerical integration of the equation or via simulations of discrete growth models within the same universality class; see, for instance, Refs.~\cite{Barabasi1995,Takeuchi2018,Oliveira2022} and other therein.

In general, the nonlinear behavior of the KPZ equation is governed by the coupling parameter $g=\lambda^2 D/\nu^3$ \cite{Amar1990,Hentschel1991,Barabasi1995} and hence becomes stronger when $\lambda$ and/or $D$ increase, or when $\nu$ decreases. The magnitude of $g$ correlates with the numerical stability of schemes employed to simulate the equation, which deteriorates as the strength of the nonlinearity increases \cite{Gallego2011}. Moreover, the KPZ equation exhibits a non-equilibrium roughening transition (RT) as a function of $g$, separating a smooth phase from a rough phase \cite{Tang1990,Barabasi1995,Krug1997,Tauber2014}. For $d<2$, the system is always in the rough phase and displays the characteristic features of the KPZ universality class. In contrast, for $d>2$, the rough phase can only be reached when the coupling parameter exceeds a critical value, $g>g_c$. When $g<g_c$, the nonlinear term is irrelevant and the scaling behavior corresponds to that of the EW universality class in the corresponding dimension, a regime known as weak coupling. When $g>g_c$, the nonlinear term becomes relevant and KPZ scaling emerges. The critical value $g_c$ increases with the system dimension, with $d=2$ being the lower critical dimension for this transition.

The scaling exponents of the KPZ universality class are expected to depend on the spatial dimension $d$ provided that $d<d_u$, as is the case for equilibrium critical systems \cite{Kardar2007,Tauber2014}. Indeed, there exists a substantial body of numerical and theoretical work aimed at determining the KPZ critical exponents as functions of the system dimension, see, for example, Refs.~\cite{Barabasi1995,Kelling2011,Kelling2016,Alves2014,Oliveira2022} and other therein. However, the existence and precise value of the upper critical dimension $d_u$ for KPZ remain unresolved. Analytically, different predictions have been put forward: some studies suggest that $d_u\le4$ \cite{Halpin-Healy1990,Bouchaud1993,Doherty1994,Colaiori2001}, others propose that $d_u>4$ \cite{Kloss2012,Kloss2014,Kloss2014-2,Canet2025}, and other argue for $d_u=\infty$ \cite{Castellano1998,Castellano1998-2}. Numerical simulations of several models within the KPZ class provide strong evidence that, if finite, $d_u$ must be large, as it has not been observed in simulations reaching dimensions as high as $d=15$ \cite{Kim2014,Alves2014,Alves2016,Oliveira2022}. A summary of the growth exponent values obtained from simulations up to $d=15$ can be found in Ref.~\cite{Oliveira2022}, while further discussion and more recent numerical estimates, in particular for the $3+1$-dimensional KPZ/DPRM growth exponent, are provided in Ref.~\cite{HalpinHealy2025}. Additional discussion of the KPZ behavior in the infinite-dimensional limit can be found in the final section of Ref.~\cite{HalpinHealy2015}.

For the tensionless KPZ equation, significantly less is known about its properties due to its being marginal to morphological instabilities, which makes its numerical integration particularly difficult. Note that, from the point of view of the KPZ equation, $g\to\infty$ in the $\nu\to 0$ TKPZ limit so that nonperturbative effects are expected in the TKPZ scaling behavior. Recent advances have enabled reliable simulations \cite{Cartes2022,RodriguezFernandez2022}, revealing that the TKPZ equation defines a universality class distinct from the standard KPZ and exhibits intrinsically anomalous scaling \cite{Lopez1997}. In one space dimension, the critical exponents are $\alpha=z=\beta=1$\cite{RodriguezFernandez2022}. 
A tensionless fixed point with $z=1$ has also been identified through RG analysis and is expected to hold for all $d>1$ with dimension-independent exponents \cite{Fontaine2023,Gosteva2024}. Incidentally, a non-local generalization of the KPZ equation has been identified \cite{Nicoli2009b} which features the same set of dimension-independent exponents but satisfies a standard dynamic scaling ansatz \cite{Nicoli2011,Nicoli2013}, as experimentally assessed for $d=2$ \cite{Castro2012}.

Prior to our work, several attempts had been made to explore the infinite-dimensional limit of the EW, KPZ, and/or TKPZ equations through simulations of discrete models, and also by integrating the equations explicitly, on the Bethe lattice or, more precisely, on its approximation by finite Cayley trees \cite{Saberi_2013,Oliveira2021,Marcos2025}. However, a central conclusion of these studies is that such type of substrates is dominated by boundary effects and is therefore not well suited for probing the high-dimensional limit. In our present paper, we instead aim to investigate this limit using a complete graph as the underlying substrate, which provides a standard mean-field representation of an effectively infinite-dimensional system \cite{Nishimori2001,Parisi2020}. Moreover, since all vertices are topologically equivalent, the complete graph has no distinguished boundary and thus avoids the boundary effects inherent to finite Cayley trees \cite{Oliveira2021,Marcos2025}.

In this study, we perform numerical simulations of the EW, KPZ, and TKPZ equations on complete graphs. Our objective is to further investigate the nature of the upper critical dimension in the KPZ universality class. To this end, we systematically tune the nonlinear coupling parameter $\lambda$, ranging from the Gaussian regime corresponding to the EW limit ($\lambda = 0$) to the strongly nonlinear case effectively realized when surface tension is suppressed ($\nu = 0$), as in the TKPZ model. Within this framework, we derive theoretical results for the EW equation that are in excellent agreement with the numerical simulations, investigate the stability of the numerical integration of the KPZ equation comparing its behavior with the corresponding EW predictions, and show that, for the TKPZ equation, the stabilization required for numerical integration prevents direct access to the true underlying dynamics.

This paper is organized as follows. In Section~\ref{observables}, we outline the numerical integration scheme employed and introduce the observables analyzed throughout the study. Section~\ref{sec:results} is devoted to the presentation of the numerical results for the EW, KPZ and TKPZ equations. Concluding remarks and final considerations are provided in Section~\ref{sec:concl}, while a number of additional technical aspects are reported in several appendices at the end.

\section{Definitions and details} \label{observables}

In this section, we introduce the concept of a complete graph, describe the integration scheme used for the continuous equations, and specify the observables measured in our simulations.

\subsection{Complete graph}

A complete graph, also known as a fully connected graph, is a simple undirected graph in which every pair of distinct vertices is connected by an edge. As a consequence, all vertices have the same degree, equal to $N - 1$, where $N$ is the total number of vertices in the graph. Unlike tree-like structures such as the Cayley tree, a complete graph is maximally connected and contains the largest possible number of edges compatible with the number of nodes, namely $N(N - 1)/2.$

In contrast to sparse graphs or hierarchical structures, no notion of shells, layers, or distance from a root can be naturally defined, since every node is directly connected to all other nodes. In fact, the diameter of a complete graph is equal to one, as any vertex can be reached from any other through a single edge. Because all sites are topologically equivalent, no distinction can be drawn between surface and bulk nodes, and boundary effects are therefore absent by construction. This strong topological homogeneity makes the complete graph an ideal framework for testing mean-field approaches, since each node experiences the same effective environment determined by the global state of the system~\cite{Nishimori2001,Parisi2020}.

Consequently, many statistical physics models defined on complete graphs reproduce the results of their mean-field formulations. Paradigmatic examples of this correspondence include the Ising model, which in this context is known as the Curie–Weiss model and can be solved analytically \cite{Baxter2016}, and which also serves as a standard benchmark for simple numerical simulations \cite{Newman1999}. Contact processes have also been extensively studied on complete graphs, where their behavior has been shown to coincide with mean-field predictions \cite{Peterson2011,Xue2017}, as well as in the context of epidemic spreading models \cite{OttinoLffler2017,Guo2013}. In addition, many other models have been investigated on complete graphs, including voter models \cite{Lipowski2022,Azhari2022,Fronczak2017,Sood2008} and bond percolation problems \cite{Huang2018}, among others.

\subsection{Integration scheme}

In this subsection, we present the discretization scheme employed for the numerical integration of the governing differential equations. This requires specifying the formulation of the Laplacian and gradient operators on networks with arbitrary topology, where no natural notion of direction exists.

In Ref.~\cite{Marcos2025}, several discretizations of the Laplacian and the squared gradient on a network are introduced and discussed. In addition, their numerical stability is analyzed. In particular, the standard discretization of these operators is given by
\begin{eqnarray}
\label{eq:discretizacion}
\nabla^2 h(x_i,t) &=& \sum_{j\sim i} h_{j}-\mathrm{deg}(i)h_{i}, \nonumber \\
(\nabla h)^2(x_i,t) &=& \sum_{j\sim i} (h_j-h_i)^2,
\end{eqnarray}
where $\sum_{j \sim i}$ denotes the sum over all nodes $j$ that are neighbors of node $i$, $\mathrm{deg}(i)$ denotes the degree of node $i$ and $h_i=h(x_i,t)$. In our case, due to the special topology of a complete graph, these formulas can be simplified as follows:
\begin{eqnarray}
\label{eq:discretizacion2}
\nabla^2 h(x_i,t) &=& \sum_{j\ne i} (h_{j}-h_{i}), \nonumber \\
(\nabla h)^2(x_i,t) &=& \sum_{j\ne i} (h_j-h_i)^2.
\end{eqnarray}
Thus, the discretized KPZ equation on a complete graph would read
\begin{eqnarray}
\label{eq:integracion}
h_{i}^{n+1} &=& h_{i}^n + \nu\Delta t\sum_{j\ne i} \left(h_{j}^n - h_{i}^n\right) \nonumber\\
    & & + \frac{\lambda\Delta t}{2}\sum_{j\ne i} \left(h_j^n - h_i^n\right)^2
    + \sqrt{2D\Delta t}\hspace{1mm}\eta_i^n \, .\nonumber\\
    &&t=n\Delta t.
\end{eqnarray}
where $\eta_i^n$ denotes a normally distributed random variable with zero mean and unit variance, typically generated using the standard Box--Muller algorithm \cite{Box1958}. Since all the $h$ terms on the right-hand side of the equation correspond to the $n$-th time step, the method is explicit. Hereafter, we will refer to this integration method as the standard (ST) scheme.

This discretized equation can be rewritten in terms of the deviations from the mean, $u_i = h_i - \bar{h}$, where $\bar{h}=\frac{1}{N}\sum_i h_i$, yielding (see Appendix \ref{appendix_ui})
\begin{eqnarray}
\label{eq:integracion_ui}
u_i^{n+1} &=& u_i^n
- \nu \Delta t\, N u_i^n
+ \frac{\lambda \Delta t\, N}{2}
\left[ (u_i^n)^2 - w^2 \right] \nonumber\\
&+& \sqrt{2D\Delta t}\,
\left( \eta_i^n - \xi^{\,n} \right),
\end{eqnarray}
where $w^2 = \frac{1}{N}\sum_j (u_j^n)^2$ denotes the roughness at time step $n$ and for that specific noise realization and $\xi^{\,n}=\frac{1}{N}\sum_i \eta_i^n$ denotes the spatial average of the noise at time step $n$. In this equation it is clear that the nonlinear term introduces a quadratic contribution that can lead to numerical instabilities. For sufficiently large fluctuations, that is, for $u_i^n$ with large magnitude, the dynamics may exhibit uncontrolled numerical growth, potentially leading to numerical overflow. Consequently, absolute stability cannot be guaranteed for arbitrary $\Delta t$, regardless of how small the time step is. Nevertheless, for sufficiently small time steps and large system sizes $N$, stability can be maintained over physically relevant simulation time ranges, as will be discussed later. It is important to remark here that an implicit method would not by itself solve the instability problems caused by the discretized nonlinear KPZ term. Moreover, a fully implicit treatment of this nonlinear term would also be computationally much more expensive.

A way to address this numerical instability issue is to replace the nonlinear term $(\nabla h_i)^2$ by a control function $f\!\left((\nabla h_i)^2\right)$, where $f(x) = (1 - e^{-c x})/c$ and $c>0$ is an adjustable parameter \cite{Marcos2025,Dasgupta1996,Dasgupta1997}. This procedure, known as the controlled-instability approach, mitigates the onset of numerical instabilities that arise from the nonlinear term. In this framework, the parameter $c$ must be chosen as small as possible in order to accurately approximate the KPZ equation, while still being sufficiently large to suppress numerical instabilities.
Accordingly, the discretized equation employed in this work to integrate the KPZ equation reads
\begin{eqnarray}
\label{eq:integracion_control}
h_{i}^{n+1} &=& h_{i}^n + \nu\Delta t\sum_{j\ne i} \left(h_{j}^n - h_{i}^n\right) \nonumber\\
    & & + \frac{\lambda\Delta t}{2}f\left(\sum_{j\ne i} (h_j^n-h_i^n)^2\right)+\sqrt{2D\Delta t}\hspace{1mm}\eta_i^n\,.
\end{eqnarray}
Hereafter, we will refer to this integration method as the control instability (CI) scheme. Two important points should be emphasized here. First, in principle the instability originates from the discretization of the KPZ equation, rather than from the equation itself. Second, this modification corresponds to the introduction of an infinite series of higher-order powers of $(\nabla h_i)^2$, with coefficients that depend on the value of $c$, which should not alter the hydrodynamic scaling behavior\cite{Dasgupta1996,Dasgupta1997,Miranda2008,Ales2019,Song2021}, provided that the value of this parameter is chosen so that the control acts only on rare, large-gradient events responsible for numerical instabilities. If the control parameter value is not chosen appropriately, or if the control acts frequently, the CI scheme can indeed alter the effective dynamics. In that case, it should be regarded primarily as a diagnostic tool rather than as a faithful regularization of the original dynamics.

\subsection{Observables}

The primary quantity analyzed in this study is the global interfacial width or roughness, denoted by $w(t)$. It is defined through the second moment of height fluctuations as \footnote{
In our notation, $w^2$ denotes the spatial average, $w^2=\overline{u_j^2}$, whereas $w^2(t)$ also includes the average over noise realizations, $w^2(t)=\left\langle \overline{u_j^2} \right\rangle$.}
\begin{equation}
	\label{eq:width}
	w^2(t)=\left\langle \overline{\left[h_{i}(t)-\bar{h}(t)\right]^2} \right\rangle,
\end{equation}
where the overline indicates an average over spatial coordinates, $\overline{(\cdots)} \equiv (1/N)\sum_i (\cdots)$, and the angular brackets $\langle{\cdots}\rangle$ represent the average taken across independent realizations of the stochastic noise.

Since all nodes in a complete graph are at distance one from each other, it is not possible to meaningfully define spatial correlations. In fact, standard quantities in the analysis of kinetic roughening, such as the height-difference correlation function, become simply proportional to the interface width. However, it is still possible to measure temporal correlations and investigate the aging properties of growing fronts. In particular, one may evaluate the two-time height autocorrelation function \cite{Takeuchi2012,Takeuchi2018,Henkel2012,Odor2014},
\begin{equation}
\label{eq:aging}
C_t(t,t_0)=\langle h_i(t)h_i(t_0)\rangle - \langle h_i(t) \rangle \langle h_i(t_0) \rangle ,
\end{equation}
where $t_0$ is an earlier (waiting) time, $t>t_0$ is the observation time and in principle there can be an implicit dependence on the particular choice of the node $i$. Since all nodes are statistically equivalent, a space average can be performed in order to improve the statistical accuracy. It is more natural to express this observable in terms of the local fluctuations $u_i$. In the thermodynamic limit, the autocorrelation function can then be rewritten as follows (see Appendix~\ref{appendix_aging_formula}),
\begin{equation}
\label{eq:aging_3}
C_t(t,t_0)=\langle u_i(t)u_i(t_0)\rangle.
\end{equation}
This observable provides direct insight into memory effects~\cite{Takeuchi2012,Henkel2012,Odor2014,DeNardis2017} and is commonly associated with \emph{aging} phenomena. In this context, \emph{aging}, also referred to as \emph{physical aging}~\cite{Odor2014}, refers to situations in which a system (i) undergoes a slow, non-exponential relaxation toward its stationary state(s), (ii) lacks time-translation invariance, and (iii) exhibits dynamical scaling~\cite{Henkel2012}. In particular, one expects this observable to scale as
\begin{equation}
    \label{eq:escalado_aging}
    C_t(t,t_0)\sim t_0^{-b} f_C(t/t_0),
\end{equation}
where the scaling function $f_C$ is expected to behave as $f_C(y)\sim y^{-\rho}$ as $y\rightarrow\infty$~\cite{Henkel2012,Rthlein2006}. The fact that $C_t(t,t_0) \to 0$ as $t/t_0 \to \infty$ means that, although fluctuations increase with time, their temporal correlations eventually vanish and the system ``forgets'' its previous conditions. This occurs for EW interfaces~\cite{Rthlein2006} and for flat KPZ interfaces~\cite{Takeuchi2012}. In contrast, circular KPZ interfaces exhibit a non-zero saturation value of the time correlation $ C_t $, indicating the presence of an \emph{infinite-time memory}, i.e., trajectories do not fully decorrelate, even asymptotically~\cite{Takeuchi2012}.

When the interface reaches a stationary regime, aging disappears; one finds \begin{equation}
    C_t(t,t_0) \simeq C_{\mathrm{stat}}(t - t_0),
\end{equation}
showing that correlations depend only on time differences, i.e., time-translation invariance is restored. Furthermore, above the upper critical dimension, aging ---understood as a scale-free dependence on the ratio $ t/t_0 $--- is expected to be essentially absent, since correlations rapidly converge to a stationary form. This is the case for the EW equation, which exhibits scaling with $t/t_0$ in one dimension, logarithmic dependence at the critical dimension $d=2$, and the recovery of time-translation invariance in dimensions above the critical one \cite{Rthlein2006}.

In addition to temporal correlations, several works have studied the time power spectrum of the interface \cite{Lauritsen1993,Takeuchi2017,VaquerodelPino2025} in order to extract information about the critical exponents of the system. It can be computed as \cite{Lauritsen1993}
\begin{equation}
\label{eq:power_spectra}
    S(\omega)=\frac{1}{T}\left\langle \left| \sum_{t=0}^{T-1}\bar{h}(t)e^{i\omega t} \right|^2\right\rangle.
\end{equation}
For regular lattices and below the upper critical dimension, where the interface exhibits FV scaling, the time power spectrum scales as
\begin{equation}
    S(\omega)\sim V^{-1}\omega^{-\psi},
\end{equation}
where $V$ is the volume of the system ($V=L^d$ for regular lattices), and the exponent $\psi$ is given by \cite{Lauritsen1993}
\begin{equation}
\label{eq:psi}
    \psi=1+\frac{2\alpha+d}{z}.
\end{equation}

Additionally, recent advances in the study of surface kinetic roughening, especially concerning KPZ dynamics \cite{Kriecherbauer2010,HalpinHealy2015,Takeuchi2018}, have revealed that universality encompasses more than just the values of scaling exponents. In particular, it has been shown that appropriately rescaled height fluctuations exhibit universal statistical properties. Defining the normalized fluctuations with respect to the instantaneous interface width as
\begin{equation}
	\label{eq:chi}
	\chi_i(t) = \frac{u_i(t)}{w(t)} ,
\end{equation}
the associated probability density function (PDF) of the random variable $\chi$ attains a stationary form throughout the growth regime and is common to all models belonging to the same universality class \cite{Kriecherbauer2010,HalpinHealy2015,Carrasco2016,Takeuchi2018,Carrasco2019}. Notably, in one spatial dimension ($d=1$), height fluctuations in the KPZ universality class are governed by one of the well-known Tracy--Widom (TW) distributions, with the specific form determined by the fact that the system size remains time-independent or else evolves with time \cite{Kriecherbauer2010,HalpinHealy2015,Takeuchi2018}, whereas those in the EW universality class are Gaussian and they differ from both, TW and Gaussian forms, for the TKPZ equation \cite{RodriguezFernandez2022}. By contrast, for other substrates, such as a complete graph, the nature of the fluctuation distribution remains an open issue. Above the upper critical dimension, the form of the height-fluctuation distribution is, in general, not well established and may depend on the specific model under consideration. Numerical solutions of the EW equation, for example, indicate that the kurtosis shifts toward increasingly negative values as the substrate dimension grows from $d=3$ to $d=5$. In contrast, within the same range of dimensions, the discrete Random Deposition with Surface Relaxation (RDSR) model displays a rising kurtosis that remains positive. This occurs even though both models belong to the EW universality class and generate interfaces that are nearly flat~\cite{Alves2014,Marcos2026}. The discrepancy can be attributed to the distinct microscopic dynamics of the two systems: the EW equation evolves through smooth continuum dynamics, whereas the discrete RDSR model permits only a restricted set of height configurations. These results highlights that, above the upper critical dimension, microscopic features can influence quantities such as the height-fluctuation distribution and its higher-order cumulants in a non-universal way.

Moreover, we have also evaluated the skewness and the kurtosis of these height distributions in order to further characterize their statistical properties. The skewness $S$ and the kurtosis $K$ are defined, respectively, in terms of the deviations from the mean $u_i$ as
\begin{equation}
	S = \frac{\langle u_i^3 \rangle}{\langle u_i^2 \rangle^{3/2}}, 
	\qquad
	K = \frac{\langle u_i^4 \rangle}{\langle u_i^2 \rangle^{2}}.
\end{equation}

The uncertainties of all the observables computed in our simulations were estimated using the jackknife resampling technique \cite{Young2015,Efron1982}. In addition, temporal data were accumulated into logarithmically spaced bins, within which the corresponding averages were computed. Throughout this work, a total of 100 such time windows was employed. A detailed discussion of both the jackknife methodology and the time-binning procedure can be found in Appendix B of Ref.\ \cite{Barreales2020}.

\section{Results} \label{sec:results}

In this section, we present the results of the numerical integration of the EW, KPZ, and TKPZ stochastic differential equations on a complete graph. We also analyze the numerical stability of the employed integration schemes.

\subsection{EW equation}

\begin{figure}[!t]
\centering
\includegraphics[width=0.6\textwidth]{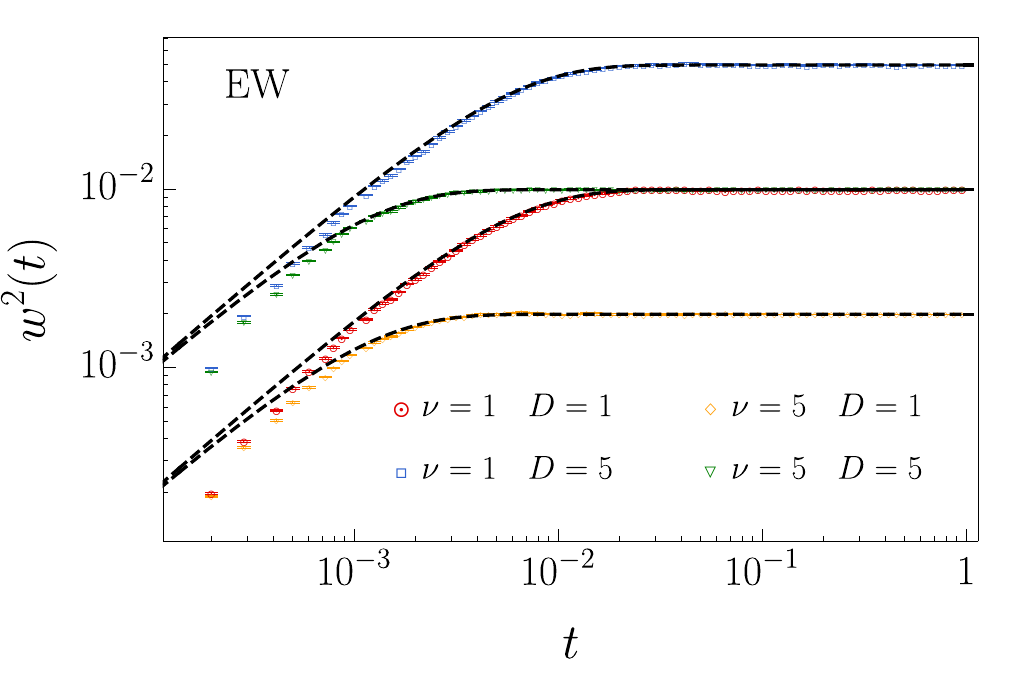}
\caption{Log-log plot of the squared roughness $w^2(t)$ as a function of time $t$ for the EW equation. Data correspond to different combinations of the parameters $\nu$ and $D$ (see legend). The system size is $N=100$ in all cases. Dashed black lines indicate the theoretical prediction given by Eq.~\eqref{eq:rugosidad_EW}.}
\label{fig:w_EW}
\end{figure}

Due to the linearity of the EW equation and the simplicity of the topology of a fully connected graph, it is in fact possible to derive analytically a closed equation for the front roughness as a function of time. In Appendix~\ref{appendix_EW_w}, we detail the calculations required to obtain that
\begin{equation}
\label{eq:rugosidad_EW}
w^2(t)=\frac{D (N-1)}{\nu N^2}\left( 1 - e^{-2 \nu N t} \right).
\end{equation}
Figure~\ref{fig:w_EW} shows the time evolution of the square of the global roughness, $w^2(t)$, for four representative sets of parameter values. The agreement with the theoretical prediction given by Eq.~\eqref{eq:rugosidad_EW} is excellent. The only region where this agreement deteriorates is at short times. We have verified that this discrepancy is related to the temporal discretization of the differential equation. By using smaller values of $\Delta t$, the theoretical prediction can be reproduced down to arbitrarily short times.

A direct implication of this result is that, in the thermodynamic limit ($N \to \infty$), the system exhibits a flat surface, i.e., $w(t) = 0$, in agreement with the behavior expected for the EW equation above its upper critical dimension $d_u = 2$.

\begin{figure}[!t]
\centering
\includegraphics[width=0.6\textwidth]{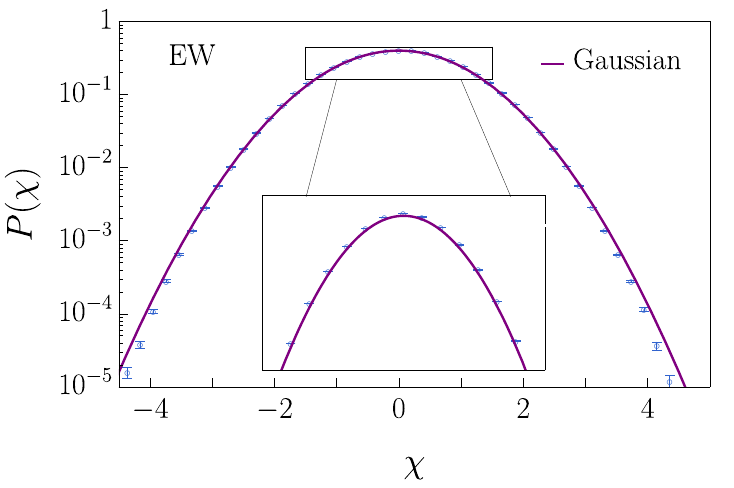}
\caption{Histogram of the rescaled height fluctuations $\chi$ [see Eq.~\eqref{eq:chi}] for the EW equation $N=100$, $\nu=1$ and $D=1$. The inset shows a magnification of the central region in the interval $-1.5<\chi<1.5$. The solid purple line corresponds to a Gaussian distribution.}
\label{fig:chi_EW}
\end{figure}

Furthermore, we have confirmed that the rescaled height fluctuations in the EW equation follow a Gaussian distribution. Figure~\ref{fig:chi_EW} shows the fluctuation PDF from simulations of the EW equation, which is seen to perfectly fit a Gaussian. This behavior is obtained in all our simulations of the EW equation. 

\begin{figure}[!t]
\centering
\includegraphics[width=0.6\textwidth]{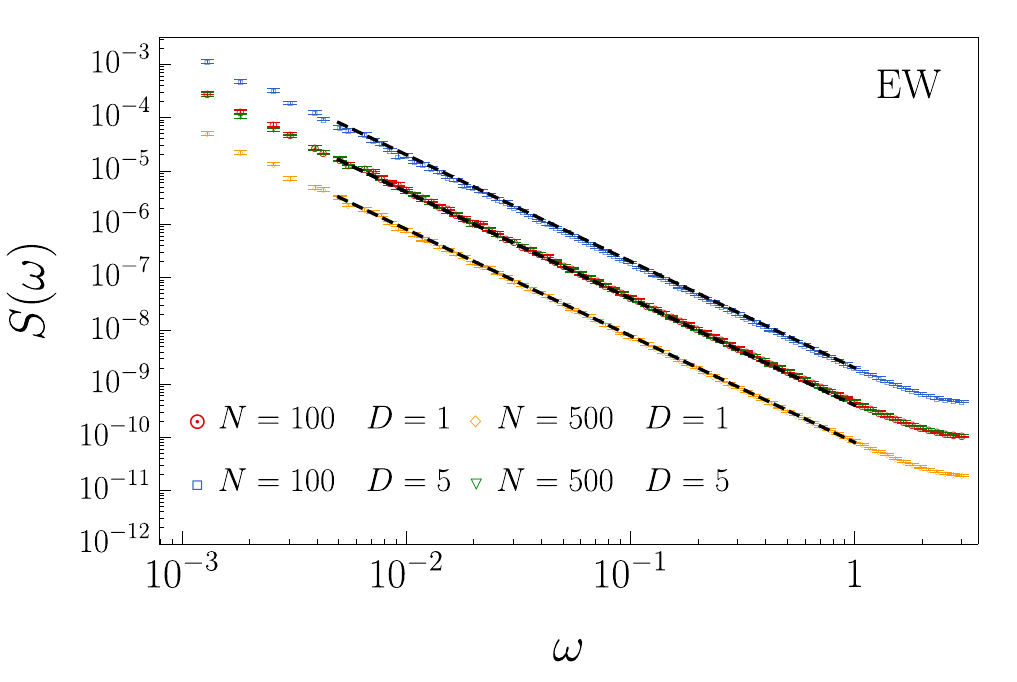}
\caption{Time power spectrum $S(\omega)$ [see Eq.~\eqref{eq:power_spectra}] for the EW equation. Data are shown for different values of $N$ and $D$ (see legend), with $\nu=1$. Dashed black lines indicate the theoretical prediction given by Eq.~\eqref{eq:ew_ps_num2}; see Eq.~\eqref{eq:ew_ps_num} in Appendix~\ref{appendix_EW_ps} for details on the prefactor.}
\label{fig:ps_EW}
\end{figure}

Figure~\ref{fig:ps_EW} shows the time power spectra, $S(\omega)$, for four representative sets of parameter values of the EW equation. The slope observed in the figure follows $S(\omega)\sim \omega^{-2}$ quite accurately. Indeed, as in the case of the roughness, owing to the linear nature of the EW equation and the simple topology of a fully connected graph, this dependence can be obtained analytically. The detailed calculations are presented in Appendix~\ref{appendix_EW_ps} and confirm that the relation 
\begin{equation}
\label{eq:ew_ps_num2}
S(\omega)\propto\frac{D}{N\omega^2}
\end{equation}
is exact for the EW equation on a complete graph. The proportionality constant depends on the Fourier-transform convention and the normalization used. Incidentally, the implied $\psi=2$ value coincides with the behavior obtained by inserting the EW universality class exponents, Eq.~\eqref{eq:exponentes_ew}, into the expected form for $\psi$, Eq.~\eqref{eq:psi}, and corresponds to kinetic roughening systems in which the hyperscaling scaling relation, $2\alpha+d=z$ holds, such as those governed by a linear height equation (such as EW itself), or by one featuring conserved dynamics and non-conserved noise \cite{Lauritsen1993}.

\begin{figure}[!t]
\centering
\includegraphics[width=0.6\textwidth]{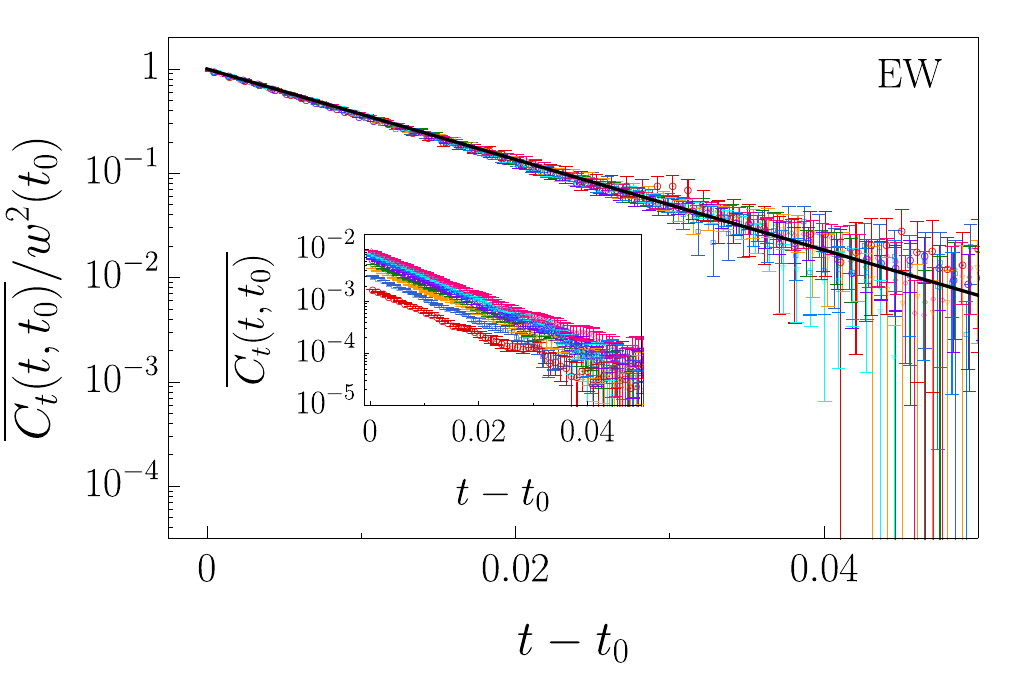}
\caption{Spatially averaged two-time height autocorrelation function, rescaled by the squared roughness, $\overline{C_t(t,t_0)}/w^2(t_0)$, as a function of the time difference $t-t_0$ for the EW equation, shown for different waiting times $t_0=\{10^{-3},2\cdot10^{-3},3\cdot10^{-3},4\cdot10^{-3},6\cdot10^{-3},8\cdot10^{-3},10^{-2},5\cdot10^{-2}\}$, which appear bottom to top in the inset. Parameters are $N=100$, $\nu=1$, and $D=1$. The solid black line corresponds to the theoretical prediction given by Eq.~\eqref{eq:aging_pred_EW}.  Inset: Same data shown without rescaling by the squared roughness $w^2(t_0)$.}
\label{fig:aging_EW}
\end{figure}

Finally, for the EW dynamics the spatially averaged two-time height autocorrelation function $\overline{C_t(t,t_0)}$ admits a closed analytical expression. A direct calculation (see Appendix~\ref{appendix_aging_EW}) yields
\begin{equation}
    \label{eq:aging_pred_EW}
    \overline{C_t(t,t_0)} = w^2(t_0)\, e^{-\nu N (t-t_0)}.
\end{equation}
Figure~\ref{fig:aging_EW} shows the spatially averaged two-time height autocorrelation function, rescaled by the squared roughness, $\overline{C_t(t,t_0)}/w^2(t_0)$, as a function of the time difference $t-t_0$, for a representative set of parameters of the EW equation and several waiting times $t_0$, while the inset shows the same data without rescaling by the roughness $w^2(t_0)$. The agreement with the theoretical prediction given by Eq.~\eqref{eq:aging_pred_EW} is excellent.

This aging behavior in the EW dynamics on a complete graph stems from the presence of a single relaxation scale in the fluctuations: once the global mode is removed, the local deviations $u_i=h_i-\bar h$ relax toward stationarity with rate $\nu N$, see Eq.~\eqref{eq:integracion_ui}. Accordingly, the roughness saturates exponentially and the two-time correlations decay on a finite, $N$-dependent time scale, implying rapid loss of memory of the initial configuration. 

More precisely, the typical relaxation time controlling the growth-to-stationary crossover of the width is $\tau = 1/(2\nu N)$. For waiting times $t_0\gtrsim \tau$, the roughness saturates and the two-time height autocorrelation functions collapse onto a single curve. Moreover, temporal correlations rapidly become effectively time-translation invariant, $C_t(t,t_0)\simeq C_{\mathrm{stat}}(t-t_0)$, and for fixed $t_0$ the two-time autocorrelation decays to zero as $t-t_0\to\infty$, i.e., trajectories decorrelate on a finite time scale. Additionally, no nontrivial aging collapse as a function of the ratio $t/t_0$ is expected: the dependence on $t$ and $t_0$ is not governed by $t/t_0$ but by the dimensionless product $\nu N (t-t_0)$ and by the amplitude $w^2(t_0)$, which saturates on the time scale $\tau$.

In this sense the complete-graph EW process exhibits, at most, \emph{trivial aging} during the short preasymptotic regime $t_0\lesssim\tau$ [through the $t_0$-dependence of the prefactor $w^2(t_0)$], while the scale-free aging typical of low-dimensional substrates is essentially absent. This is fully consistent with the mean-field character of the complete graph, where correlations rapidly converge to a stationary form. Equation~\eqref{eq:aging_pred_EW} makes explicit both the finite decorrelation time $1/(\nu N)$ (loss of memory) and the absence of a scaling form in terms of $t/t_0$.

\subsection{KPZ equation}

In this subsection, we present the results of the numerical integration of the KPZ equation and a discussion of the numerical stability of the integration schemes. 

\subsubsection{Numerical stability}

Beginning with the latter issue, and as already stated in the previous section, the ST integration scheme, computed through Eq.~\eqref{eq:integracion}, is unstable for any choice of $\Delta t$, and the probability of encountering numerical overflow never vanishes. Therefore, absolute stability cannot be achieved by tuning the time step $\Delta t$, irrespective of how small it is chosen. Nevertheless, for sufficiently small time steps and large system sizes $N$, stability can be maintained over physically relevant simulation time scales and it is possible to perform the integration using the ST scheme instead of the CI scheme.

\begin{figure*}[!t]
\centering
\includegraphics[width=0.6\textwidth]{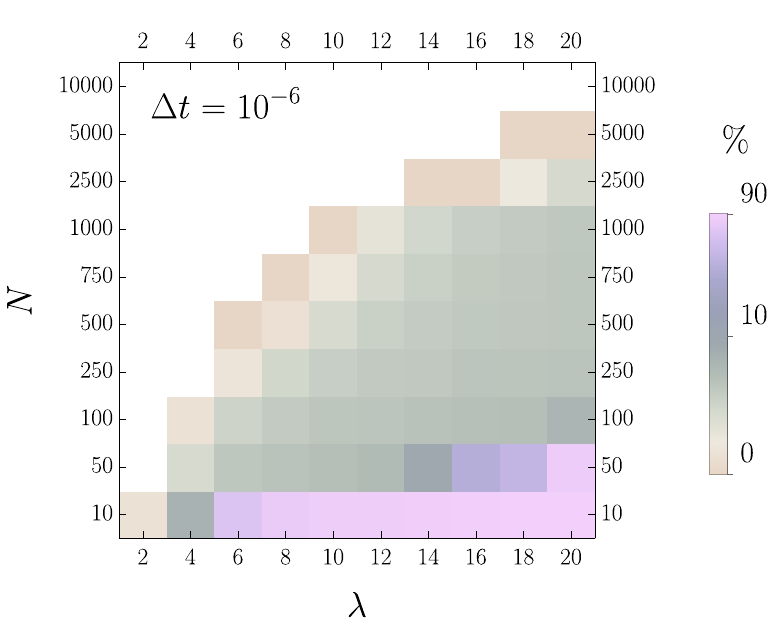}
\hspace{10mm}
\includegraphics[width=0.6\textwidth]{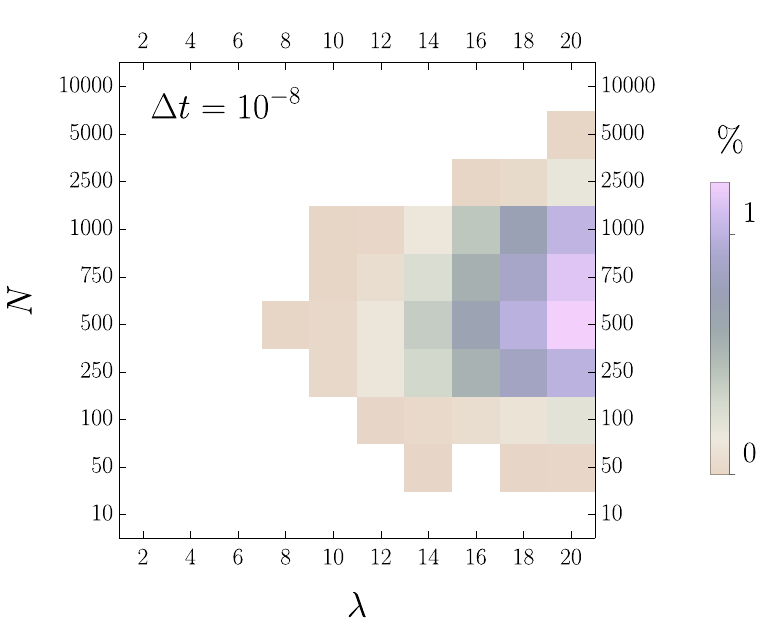}
\caption{Numerical stability of simulations of the KPZ equation using the CI scheme, for several values of $\lambda$ and system sizes $N$, using time steps $\Delta t$ as indicated in the legends. Parameters are $\nu=1$, $D=1$, $N_{\mathrm{steps}}=10^6$, and $c=0.01$. Each grid cell shows the percentage of time steps for which the gradient defined in Eq.~\eqref{eq:discretizacion2} exceeds $1/c$ and therefore the control function effectively caps this gradient. Uncolored cells correspond to $0\%$.}

\label{fig:estabilidad}
\end{figure*}

Figure~\ref{fig:estabilidad} shows a summary of the numerical stability of the simulations of the KPZ equation. In this figure, where the CI integration scheme is employed, we plot the percentage of times that the gradient, given by Eq.~\eqref{eq:discretizacion2}, exceeds the value $1/c$ and is therefore effectively controlled by the control function $f(x)$. Several conclusions can be drawn from this figure. First, the stability of the integration increases significantly as the integration time step $\Delta t$ is reduced, even though very small time steps are still required. Second, for fixed integration time step $\Delta t$ and nonlinearity $\lambda$ (or, equivalently, fixed coupling $g$), increasing the system size $N$ stabilizes the integration. Finally, for parameter choices such that the gradient never exceeds the threshold $1/c$, no control is required, and the ST integration scheme should be usable without numerical overflow.

We have verified that, for parameter choices where the latter condition is satisfied (for instance, $\lambda = 6$, $N = 1000$, and $\Delta t = 10^{-8}$), the results obtained using both integration schemes are identical. We have also observed that, in the vicinity of the numerically unstable region, the ST integration scheme begins to produce numerical overflows. For example, for $\lambda = 10$, $N = 1000$, and $\Delta t = 10^{-8}$, 137 out of 500 runs resulted in numerical overflow, whereas for $\lambda = 20$, $N = 1000$, and $\Delta t = 10^{-8}$ all runs led to overflow. Nevertheless, for the case $\lambda = 10$, $N = 1000$, and $\Delta t = 10^{-8}$, a random sample of 100 runs that did not experience overflow produced results identical to those obtained using the CI scheme.

Finally, a somewhat counterintuitive behavior deserves comment. Although numerical stability generally improves as the system size increases, Fig.~\ref{fig:estabilidad} shows that for $\Delta t = 10^{-8}$ and small system sizes (specifically $N=10$ and $N=50$), the numerical integration exhibits almost no instability. In contrast, for larger system sizes instabilities are clearly observed, and for $\Delta t = 10^{-6}$ the instability is in fact maximal precisely at these small values of $N$. Assuming that the KPZ dynamics behaves similarly to the EW case ---which, as will be shown later, is indeed the case--- the squared roughness initially is proportional to $t$ at short times before reaching saturation, approximately at a saturation time $\tau = 1/(2\nu N)$. Therefore, the final simulation times for $\Delta t = 10^{-8}$, which are $10^{-2}$, are approximately equal to the saturation time $\tau$. In these cases, the system has not yet reached sufficiently long times for the nonlinearity to become relevant. Indeed, by analyzing the roughness growth prior to saturation, namely at times $t \ll 1/(2\nu N)$, one finds that in both the EW and KPZ cases the very early-time dynamics is governed by the Random Deposition (RD) equation [namely, Eq.\ \eqref{eq:KPZ_orig} for $\nu=\lambda=0$], corresponding to a regime characterized by
\begin{equation}
w^2(t) \sim t, \qquad \chi \sim \text{Gaussian}, \qquad \bar{h}(t) \approx 0 .
\end{equation}
This behavior is not unexpected, since such transient regimes are known to occur in both EW and KPZ dynamics on regular lattices~\cite{Barabasi1995,Chou2010,Almeida2017,AaroReis2006}. Nevertheless, it helps to clarify that this early-time growth is not governed by the corresponding universality classes, but rather by RD. Notably, this behavior arises independently of the chosen parameter values, even for large couplings.

\subsubsection{Numerical results}

\begin{figure}[!t]
\centering
\includegraphics[width=0.6\textwidth]{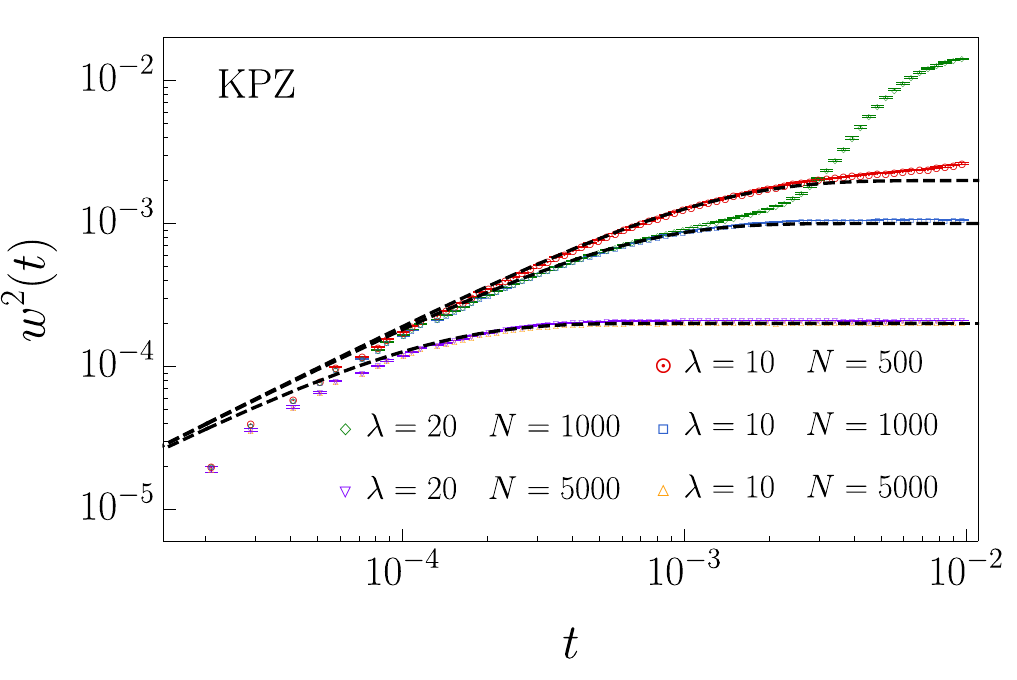}
\caption{Log-log plot of the squared roughness $w^2(t)$ as a function of time $t$ for the KPZ equation. Data correspond to different values of $\lambda$ and system sizes $N$ (see legend). Parameters are $\nu=1$, $D=1$, $\Delta t=10^{-8}$, and $c=0.01$. Dashed black lines indicate the EW prediction given by Eq.~\eqref{eq:rugosidad_EW}. The CI integration scheme is used.}
\label{fig:w_KPZ}
\end{figure}

We now proceed to discuss the results obtained for the KPZ equation. Figure~\ref{fig:w_KPZ} displays the time evolution of the squared global roughness $w^2(t)$ for the KPZ equation and several representative choices of parameters, together with the theoretical prediction of the EW equation given by Eq.~\eqref{eq:rugosidad_EW}. In this figure, the CI integration scheme was employed. As can be seen, the agreement between some of the simulation data and the theoretical prediction is very good, whereas significant deviations are observed in other cases. These deviations occur under conditions for which the control function $f(x)$ must act a large fraction of the time in order to prevent numerical overflow. By contrast, for parameter choices where the intervention of the control function is not required, the agreement is excellent. The same agreement is observed when using the ST integration scheme. For a fixed time step $\Delta t$ and a given set of equation parameters, once the system size is sufficiently large for numerical instabilities to disappear, both integration schemes yield the same results, which are in agreement with the EW theoretical prediction. Additionally, we have verified that these deviations from EW behavior, which arise under conditions where numerical instabilities are present, are not caused by a few anomalous realizations, but instead occur systematically across all simulations corresponding to that particular choice of parameters.

\begin{figure}[!t]
\centering
\includegraphics[width=0.6\textwidth]{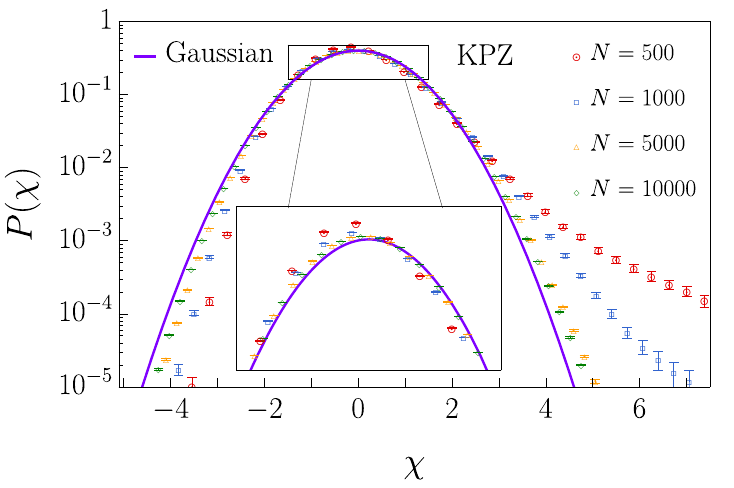}
\caption{Histogram of the rescaled height fluctuations $\chi$ [see Eq.~\eqref{eq:chi}] for the KPZ equation and several system sizes $N$ (see legend). Parameters are $\lambda=10$, $\nu=1$, $D=1$, $\Delta t=10^{-8}$, and $c=0.01$. The inset shows a magnification of the central region $-1.5<\chi<1.5$. The solid purple line corresponds to a Gaussian distribution. The CI integration scheme is used.}
\label{fig:chi_KPZ}
\end{figure}

Figure~\ref{fig:chi_KPZ} shows the rescaled height fluctuations, $\chi$, for the KPZ equation measured at times longer than the saturation time $\tau$, so that the RD transient does not influence this measurement. As can be clearly seen from the figure, these fluctuations become increasingly Gaussian as the system size increases. This trend is even more evident from the analysis of the skewness and kurtosis, which yield $S=1.2(1)$ and $K=11(2)$ for $N=500$, $S=0.39(1)$ and $K=3.7(1)$ for $N=1000$, $S=0.14(1)$ and $K=3.05(1)$ for $N=5000$, and $S=0.098(3)$ and $K=3.027(5)$ for $N=10000$. Once again, when numerical instabilities are stronger, the behavior of the fluctuations departs more significantly from the EW Gaussian form, leading to the appearance of fat tails in the $\chi$ distribution. As before, increasing the system size is sufficient to eliminate these effects. Moreover, for smaller coupling strengths $g$, the convergence toward a Gaussian distribution is achieved more easily.

It is important to note that, for certain parameter choices under which the numerical integration is stable and the ST integration scheme can therefore be safely employed, the rescaled height fluctuations $\chi$ are not strictly Gaussian. Instead, they exhibit a shape slightly reminiscent of TW distributions, with a kurtosis slightly larger than 3 and a small positive skewness, which constitutes the hallmark of nonlinear effects. However, these features disappear simply by increasing the system size. The main idea we wish to convey is that, as the system size is increased, numerical instabilities disappear first, and upon further increasing the system size, the KPZ results progressively approach those obtained for the EW equation.

\begin{figure}[!t]
\centering
\includegraphics[width=0.6\textwidth]{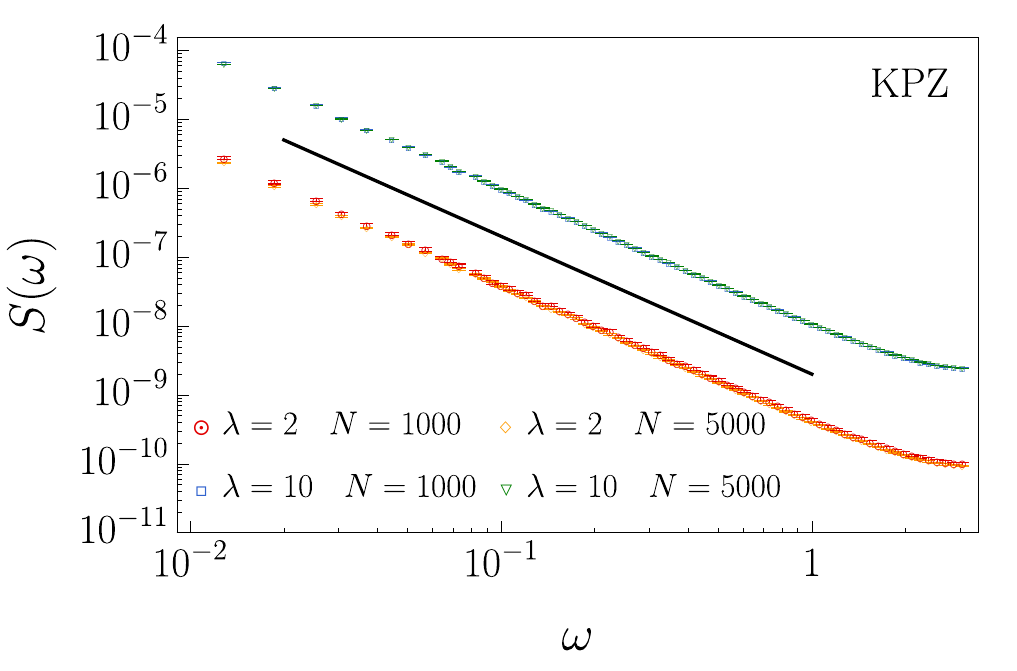}
\caption{Time power spectrum $S(\omega)$ [see Eq.~\eqref{eq:power_spectra}] for the KPZ equation. Data are shown for different values of $\lambda$ and system sizes $N$ (see legend). The remaining parameters are $\nu=1$, $D=1$, $\Delta t=10^{-8}$, and $c=0.01$. The solid black line indicates the  $S(\omega)\sim\omega^{-2}$ scaling. The CI integration scheme is used.}
\label{fig:ps_KPZ}
\end{figure}

Figure~\ref{fig:ps_KPZ} shows the time power spectra, $S(\omega)$, for four representative sets of parameter values of the KPZ equation. The slope observed in the figure follows $S(\omega) \sim \omega^{-2}$, in agreement with the behavior exhibited by the EW equation. The main difference with respect to the EW case lies in the absolute value of the spectra, which is now larger by a couple of orders of magnitude and increases with $\lambda$. Moreover, this absolute value appears to depend much more strongly on $\lambda$ than on the system size. Notably, numerical instabilities have little impact on this observable, leading to a particularly clean and robust behavior.

\begin{figure}[!t]
\centering
\includegraphics[width=0.6\textwidth]{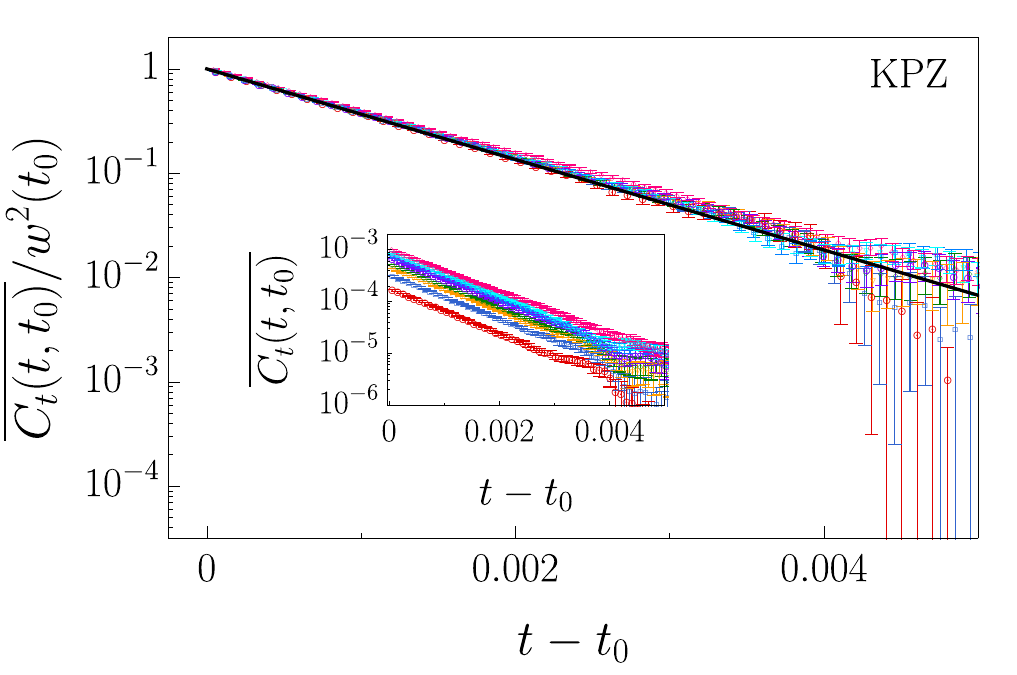}
\caption{Spatially averaged two-time height autocorrelation function, rescaled by the squared roughness, $\overline{C_t(t,t_0)}/w^2(t_0)$, as a function of the time difference $t-t_0$ for the KPZ equation, shown for different waiting times $t_0=\{10^{-4},2\cdot10^{-4},3\cdot10^{-4},4\cdot10^{-4},6\cdot10^{-4},8\cdot10^{-4},10^{-3},3\cdot10^{-3}\}$, which appear bottom to top in the inset. Parameters are $N=1000$, $\lambda=8$, $\nu=1$, $D=1$, $\Delta t=10^{-8}$, and $c=0.01$. The solid black line corresponds to the theoretical prediction given by Eq.~\eqref{eq:aging_pred_EW}. The CI integration scheme is used. Inset: Same data shown without rescaling by the squared roughness $w^2(t_0)$.
}
\label{fig:aging_KPZ}
\end{figure}

Figure~\ref{fig:aging_KPZ} shows the spatially averaged two-time height autocorrelation function, rescaled by the squared roughness, $\overline{C_t(t,t_0)}/w^2(t_0)$, as a function of the time difference $t-t_0$, for a representative set of parameters of the KPZ equation and several waiting times $t_0$, together with the theoretical prediction for the EW equation given by Eq.~\eqref{eq:aging_pred_EW}. The inset shows the same data without rescaling by the squared roughness $w^2(t_0)$. As can be seen, the agreement with the EW theoretical prediction is excellent. The aging analysis for the KPZ equation on the complete graph yields the same qualitative behavior as for the EW dynamics. In particular, only \emph{trivial aging} is observed: the scale-free aging typical of low-dimensional substrates is essentially absent, and temporal correlations rapidly converge to a stationary form. For parameter regimes in which numerical simulation of the KPZ equation becomes unstable, significant deviations from the behavior predicted by Eq.~\eqref{eq:aging_pred_EW} are observed. As in the case of the roughness, we attribute these deviations to the control function, which distorts the asymptotic scaling behavior.

One of the main conclusions drawn from our numerical integration of the KPZ equation is that, for sufficiently large system sizes, any nonlinear effects, together with the numerical instabilities associated with them, are effectively suppressed, regardless of how strong they are, and the resulting behavior coincides with that of the EW equation. Our simulations therefore suggest that, in the thermodynamic limit ($N \to \infty$) on a complete graph, the KPZ nonlinearity becomes irrelevant, leading to a flat interface whose behavior is indistinguishable from that of the EW equation. 

It is important to note that the results presented in this section also suggest that the CI integration scheme is not always helpful for numerical simulations on complete graphs. Specifically, under conditions where numerical instabilities begin to appear ---precisely where this method might be expected to be beneficial--- the control function $f(x)$ starts to distort the scaling of the expected results. This point will be further reinforced when discussing the behavior of the mean front position in the next subsection. Consequently, the CI scheme is primarily useful as a diagnostic tool for assessing the stability of the numerical integration, and only under conditions where the instability is weak can it help suppress isolated instability spikes without altering the scaling behavior. Finally, in Appendix~\ref{appendix_estabilidad}, we explain why the ST integration method remains stable and avoids numerical overflows for sufficiently small time steps and sufficiently large system sizes, despite its formal instability.

\subsection{TKPZ equation}

In this subsection, we present the results of the numerical integration of the TKPZ equation. As anticipated in the Introduction, the stabilization required to perform the integration prevents direct access to the true underlying dynamics. In practice, stable simulations can only be obtained using the CI scheme. For any combination of parameters, the use of the ST scheme ultimately leads to numerical overflow. Nevertheless, presenting these numerical results remains relevant, as they allow us to clarify and reinterpret the related observations reported in Ref.~\cite{Marcos2025}.
 
\begin{figure}[!t]
\centering
\includegraphics[width=0.6\textwidth]{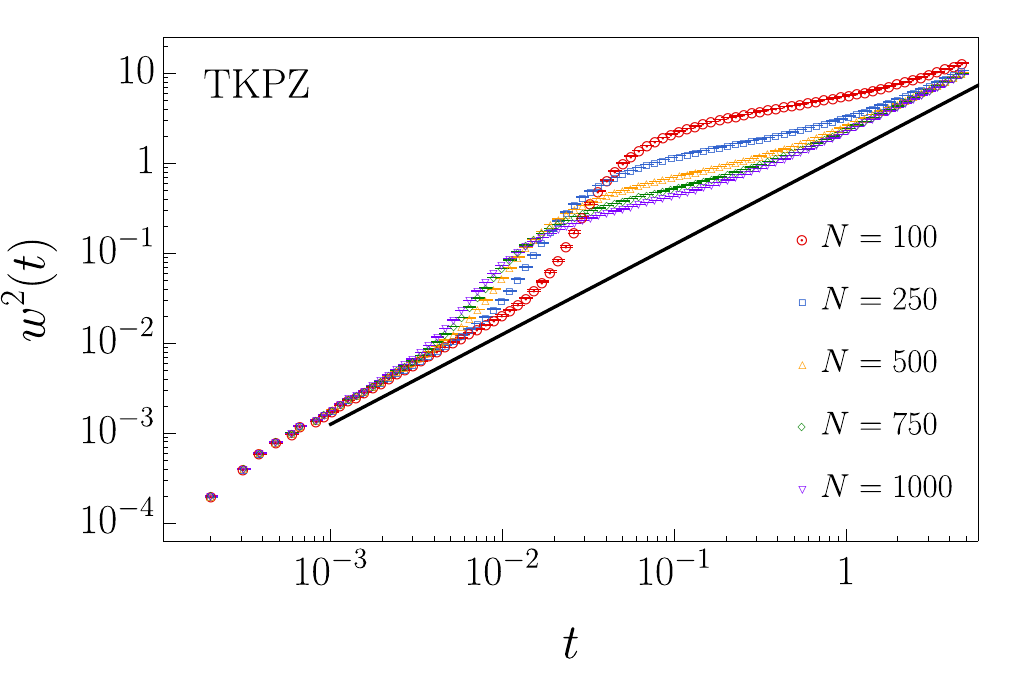}
\caption{Log-log plot of the squared roughness $w^2(t)$ as a function of time $t$ for the TKPZ equation. Parameters are $\nu=0$, $\lambda=4$, $D=1$, $\Delta t=10^{-8}$, $c=0.01$, and several system sizes $N$ (see legend). The solid black line indicates the RD scaling $w^2(t)\sim t$. The CI integration scheme is used.}
\label{fig:w_TKPZ}
\end{figure}

Figure~\ref{fig:w_TKPZ} shows the time evolution of the squared global roughness, $w^2(t)$, in the TKPZ equation, for several system sizes. The scaling observed in the figure is somewhat unusual. After a brief RD-like transient, the roughness begins to grow much more rapidly; it subsequently relaxes and returns to an RD-like growth characterized by $w^2(t)\sim t$. This behavior then persists indefinitely. RD-like behavior was previously reported for the TKPZ equation on finite Cayley trees~\cite{Marcos2025}. More remarkably, in this final regime the rescaled height fluctuations at long times follow a Gaussian distribution, in contrast to the behavior observed on Cayley trees~\cite{Marcos2025}, where a non-Gaussian distribution with positive skewness was obtained. This behavior is observed for all simulated system sizes. The only size-dependent difference is that, for larger systems, the roughness departs from the initial RD transient at earlier times.

\begin{figure}[!t]
\centering
\includegraphics[width=0.6\textwidth]{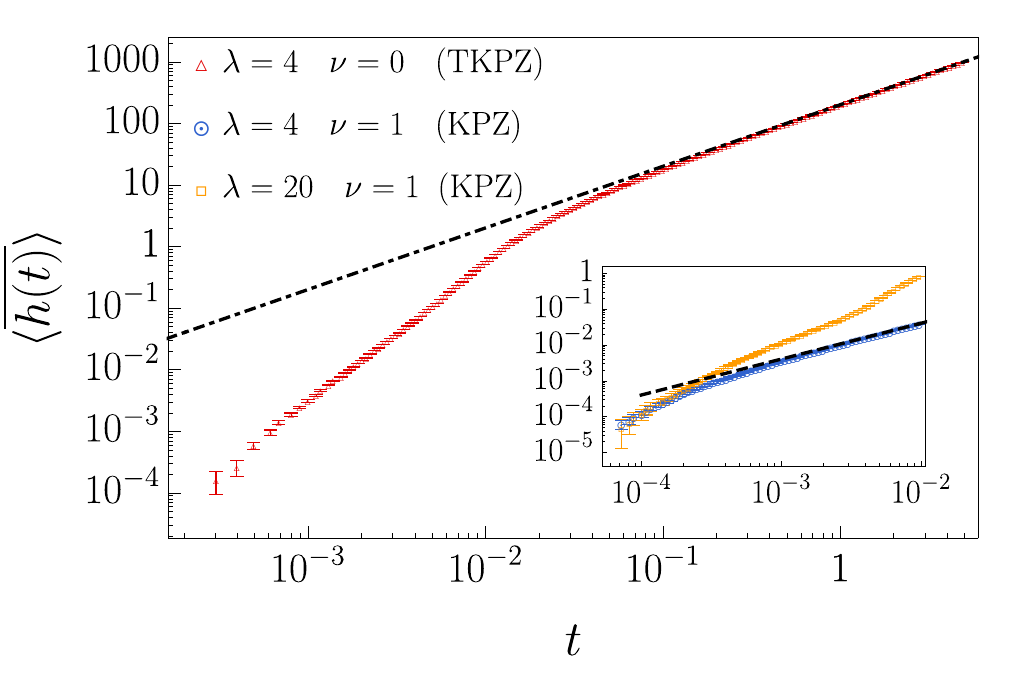}
\caption{Log-log plot of the average front position $\langle\overline{h(t)}\rangle$ as a function of time $t$ for the TKPZ equation. Parameters are $\nu=0$, $\lambda=4$, $D=1$, $N=1000$, $\Delta t=10^{-8}$, and $c=0.01$. The dot-dashed black line corresponds to the prediction $\overline{h(t)}=\lambda t/(2c)=200\,t$. The CI integration scheme is used. Inset: $\langle\overline{h(t)}\rangle$ vs $t$ for the KPZ equation with $\nu=1$, $D=1$, $N=1000$, $\Delta t=10^{-8}$, and $c=0.01$. Blue circles correspond to $\lambda=4$ and orange squares to $\lambda=20$. The dashed black line indicates the prediction $\overline{h(t)}=\lambda D t/\nu=2\,t$.}
\label{fig:h_TKPZ}
\end{figure}

We propose the following explanation for the behavior observed in the simulations of the TKPZ equation. Since no smoothening mechanism is present (as $\nu = 0$), the gradients increase rapidly and, after being processed by the control function, become effectively constant and equal to $1/c$, with $c$ the parameter of the CI scheme. In this regime, the TKPZ equation can be approximated as
\begin{equation}
    \frac{\partial h}{\partial t}=  \frac{\lambda}{2} \left(\nabla h \right)^2+ \eta(\boldsymbol{x},t) \approx \frac{\lambda}{2c}+ \eta(\boldsymbol{x},t)\equiv\eta(\boldsymbol{x},t)+F
\end{equation}
with the same right hand side as in the RD equation, plus a constant term $F=\lambda/(2c)$. This additional term modifies neither the scaling of the roughness nor the nature of the front fluctuations; it only affects the growth of the average front position, which is given by
\begin{equation}
    \label{eq:h_TKPZ}
    \langle \overline{h(t)}\rangle= F t=\frac{\lambda}{2c}t.
\end{equation}

Figure~\ref{fig:h_TKPZ} displays the time evolution of the average front position $\langle \overline{h}(t) \rangle$ for the TKPZ equation, together with the slope predicted at long times by Eq.~\eqref{eq:h_TKPZ}. The excellent agreement between the two clearly indicates that the gradient is effectively capped to a constant value by the control function. The time window over which the average front position reaches the behavior predicted by Eq.~\eqref{eq:h_TKPZ} coincides precisely with the time interval in which the roughness $w^2(t)$ exhibits the non-standard behavior. We therefore conclude that this behavior is essentially induced by the control function, which caps the gradients at different times. Once all gradients are effectively replaced by $\lambda/(2c)$, the system returns to RD-like behavior. Therefore, the true TKPZ dynamics remains numerically inaccessible, since the equation cannot be integrated without either encountering numerical overflow or trivially driving the system into an RD-like regime.

We suspect that the behavior observed on Cayley trees \cite{Marcos2025}, where the front roughness of the TKPZ equation also grows as in the RD equation, arises for essentially the same reason. In that case, the hierarchical structure of the network leads to non-Gaussian fluctuations with positive skewness, a feature typically regarded as a hallmark of the nonlinear KPZ term. However, in light of the present results, this behavior can be reinterpreted as being induced primarily by the underlying network structure rather than by the nonlinear term, which, in that substrate, is likely to be also approximated by a constant.

We emphasize that this effect appears exclusively in our simulations of the TKPZ equation. By contrast, for the EW equation the average front position is identically zero, whereas for the KPZ equation it increases linearly with time, but with a slope that is significantly smaller than that predicted by Eq.~\eqref{eq:h_TKPZ}. In fact, the time dependence of the average front position for the KPZ equation in the stationary state can be analytically approximated (see Appendix~\ref{appendix_KPZ_h}) by
\begin{equation}
\label{eq:h_KPZ}
\langle \overline{h(t)} \rangle = \frac{\lambda D}{\nu}\, t .
\end{equation}

The inset of Fig.~\ref{fig:h_TKPZ} displays the time evolution of the average front position $\langle \overline{h}(t) \rangle$ for two representative parameter choices of the KPZ equation: one for which the numerical integration is stable and another one for which it is not. In the numerically stable case, the growth rapidly converges to the behavior predicted by Eq.~\eqref{eq:h_KPZ}, whereas in the unstable case the behavior becomes non-standard, with $\langle \overline{h}(t) \rangle \not\sim t$, as an effect of the control function. Therefore, we can be confident that KPZ simulations performed under numerically stable conditions do not trivially fall into the EW universality class due to numerical artifacts, unlike the TKPZ equation, which effectively reduces to RD dynamics due to the regularization of the slopes implemented by the control function.

\section{Discussion \& Conclusions}\label{sec:concl}

In this work we have employed a complete graph as a substrate to investigate the high-dimensional limit of stochastic continuum equations for interface growth. This choice eliminates the strong boundary effects previously found on finite tree-like substrates and provides a clean setting in which all sites are topologically equivalent, allowing us to assess in a controlled way whether genuinely nonlinear KPZ physics survives in this high-dimensional approximation.

For the EW equation, linearity together with the topological homogeneity of the complete graph enables an exact closed-form expression for the global roughness, Eq.~\eqref{eq:rugosidad_EW}, which is in excellent agreement with numerical simulations. In particular, the result implies $w^2\sim N^{-1}$, so that the interface becomes asymptotically flat as $N\to\infty$. Moreover, the rescaled height fluctuations are Gaussian. The time power spectrum of the spatially averaged height displays a robust scaling $S(\omega)\sim \omega^{-2}$, seemingly valid for arbitrary $d$. In our present context this scaling is derived analytically and can be interpreted as a direct consequence of the effective Brownian dynamics of $\bar{h}(t)$ on a complete graph, see Appendix~\ref{appendix_EW_ps}. Finally, for the two-time autocorrelation function $C_t(t,t_0)$ a closed analytical expression can be obtained, in excellent agreement with the numerical results, and which clearly displays \emph{trivial aging}. In particular, the scale-free aging typical of low-dimensional substrates is essentially absent, consistent with the interpretation of the complete graph as a mean-field approximation.

For the KPZ equation, we have systematically tuned the strength of the nonlinearity and compared the resulting behavior with the EW prediction. Although numerical instabilities arise at moderate system sizes $N$ and strong couplings $g$, which distort the scaling and drive the system away from EW behavior, these discrepancies decrease systematically as $N$ increases, and the dynamics converges steadily towards that of the EW limit. This conclusion is supported by four independent observables: (i) the global roughness approaches the analytical EW form as $N$ increases, (ii) the stationary distributions of the rescaled height fluctuations $\chi$ become increasingly Gaussian with growing $N$, as quantified by skewness and kurtosis converging to their Gaussian values, (iii) the time power spectra exhibit the same $\omega^{-2}$ decay observed for the EW equation (and characteristic of a hyperscaling relation $2\alpha+z=d$, different from the $\alpha+z=2$ Galilean relation proper of the KPZ fixed point), with variations in $\lambda$ affecting only the overall amplitude and not the scaling exponent, and (iv) the two-time height autocorrelation function follows the same behavior as in the EW case and similarly exhibits no genuine aging, as correlations rapidly become effectively stationary and the scale-free aging typical of low-dimensional substrates is absent.

Taken together, these results indicate that, on a complete graph, the KPZ and EW equations are essentially indistinguishable in the thermodynamic limit: the KPZ nonlinearity becomes irrelevant as $N\to\infty$, yielding EW-like behavior with an asymptotically flat interface. Note, this correspondence turns out to be much more affected by nonuniversal aspects in the case of discrete growth models that correspond to the EW and KPZ universality classes, when defined on a complete graph \cite{Oliveira2021,Marcos2026}. For instance, the surface roughness for the random deposition with surface relaxation model (RDSR), well-known to be in the EW universality class \cite{Barabasi1995}, features an oscillatory time behavior when studied on a complete graph \cite{Marcos2026}, describing layer-by-layer growth \cite{Oliveira2021}. While the restricted solid-on-solid (RSOS) model, in the KPZ class, also features an analogous oscillating roughness on a complete graph, it shows a very different distribution of height fluctuations \cite{Marcos2026}. And in turn, the ballistic deposition model (also in the KPZ universality class) displays a roughness that increases monotonically with time (eventually reaching a time-independent value in a way which does not agree with the Family-Vicsek scaling ansatz) when defined on a complete graph \cite{Marcos2026}. The KPZ to EW correspondence in this high-dimensional setting seems definitely much more clear at the level of continuum equations, as we are presently reporting.

It is well established that the upper critical dimension of the EW equation is $d_u = 2$, i.e., the dimension at which logarithmic corrections to scaling appear both in the surface roughness and in the aging properties. For dimensions above this value, interfaces become asymptotically flat. It is also well known that a fully connected graph constitutes a standard representation of an effectively infinite-dimensional system. Therefore, the result we obtain for the EW case is consistent with these general expectations.

The result found for the KPZ equation is more interesting and considerably more subtle to interpret. We do not observe any type of logarithmic correction in any of the measured observables ---either because such corrections are genuinely absent or because they are too small to be detected within our numerical accuracy. However, this result does not allow us to conclude unambiguously on a finite or infinite value for the upper critical dimension of the KPZ universality class.\footnote{
A paradigmatic example of a system with an infinite upper critical dimension is Anderson localization \cite{AndersonRevModPhys} which, incidentally, has been recently related with the KPZ universality class \cite{Mu2024}. Another model for which a finite (if existing) upper critical dimension remains unknown is diffusion-limited aggregation (DLA) \cite{WittenSander1983,Halsey2000}; similarly, recent studies of higher-dimensional local-load-sharing (LLS) fiber-bundle models indicate that the approach to the equal-load-sharing/mean-field limit is only asymptotic, implying an infinite upper critical dimension for LLS fracture models \cite{Sinha2015,Danku2018}.}

In perturbative quantum field theory, the origin of logarithmic corrections is closely related to the merging of the Wilson–Fisher and Gaussian fixed points when the upper critical dimension is approached from below \cite{Tauber2014}. 
However, the situation in the KPZ equation is considerably more involved. Indeed, for $d>2$ the KPZ equation exhibits four fixed points: the attractive Gaussian fixed point (at $g=0$ in the coupling constant $g$ defined in Sec.\ \ref{sec:intro}), which determines the EW universality class; the attractive strong-coupling fixed point associated with the KPZ universality class; a repulsive fixed point at $g=g_c$, in between the weak (EW) and strong (KPZ) coupling regimes, associated with the non-equilibrium roughening transition (RT); and a repulsive fixed point at $g=\infty$ that corresponds to $\nu=0$ as in the TKPZ equation, termed the inviscid Burgers (IB) fixed point \cite{Canet2025}. Available analytical results indicate that the RT and KPZ fixed points move towards larger $g$ values as $d$ increases, while the IB fixed point is defined for any $d>0$ \cite{Tauber2014,Canet2025,Gosteva2024}. 

In this framework, a plausible scenario is that both fixed points, RT and KPZ, flow to infinity in the infinite-dimensional limit. Under this assumption, the entire range of values of $\lambda$ would lie within the basin of attraction of the EW fixed point, in agreement with the numerical results presented in this paper. This would imply that the upper critical dimension of the KPZ universality class is infinite. Alternative possibilities include the occurrence of this behavior (i.e., the RT fixed point reaching infinity) already at finite but very large dimensions, or instead that, at infinite dimension, the RT and KPZ fixed points attain very large yet finite values. In the latter case, our numerical simulations might fail to detect the KPZ regime simply because we have not explored sufficiently large values of $\lambda$. In any case, from this discussion there is no theoretical argument implying that logarithmic corrections must necessarily appear, regardless of whether the upper critical dimension is finite or infinite.

From a numerical standpoint, our results underscore that the choice of integration scheme plays a nontrivial role in simulations on complete graphs. Provided the dynamics remains numerically stable, typically for sufficiently large $N$ and small integration time steps $\Delta t$, the ST integration scheme yields reliable results and is fully consistent with those obtained using the CI scheme. By contrast, in parameter regimes where the nonlinear term triggers frequent activation of the control function, the CI scheme effectively alters the underlying dynamics, leading to systematic deviations in the measured observables. Under such conditions, the CI procedure should not be interpreted as a faithful regularization of the KPZ equation, but rather as a modification that suppresses numerical divergences at the cost of changing the effective evolution. Consequently, the CI scheme is best viewed as a diagnostic tool to assess the onset of numerical instabilities and to mitigate rare runaway events, whereas the extraction of asymptotic scaling behavior requires simulations to be performed in regimes where the control mechanism is essentially inactive.

Finally, for the TKPZ equation we find $w^2(t)\sim t$, together with Gaussian rescaled fluctuations at long times, indicating random-deposition-like roughening. The evidence indicates that this behavior is induced by the numerical mechanism used to control the slopes. In the absence of relaxation ($\nu=0$), gradients grow rapidly until they are effectively capped by the control function, causing the nonlinear term to approach an approximately constant drift and reducing the effective dynamics to random deposition with constant forcing. This interpretation is further supported by the linear growth of the mean height, with a slope consistent with that predicted by the corresponding effective equation. Thus, although the CI scheme stabilizes the simulations, it prevents direct access to the true TKPZ dynamics by qualitatively modifying the evolution. The main contribution of this analysis is therefore to establish this limitation and to reinterpret the related numerical observations reported in Ref.~\cite{Marcos2025}.

In contrast, for KPZ under stable integration conditions the mean height follows the expected behavior without falling into this trivial drift-dominated regime, supporting the interpretation that the effective equivalence between KPZ and EW dynamics on complete graphs originates from the properties of the substrate itself ---namely, the irrelevance of the nonlinear term as $N\to\infty$--- rather than from an unavoidable numerical artifact.

\section*{Acknowledgements}
The authors would like to thank Enzo Marinari, Giorgio Parisi, Tommaso Rizzo, and Timothy Halpin-Healy for interesting discussions and helpful comments.

\paragraph{Funding information}\sloppy This work was partially supported by Ministerio de Ciencia, Innovaci\'on y Universidades (Spain), Agencia Estatal de Investigaci\'on (AEI, Spain, 10.13039/501100011033), and European Regional Development Fund (ERDF, A way of making Europe) through Grant No.\ PID2021-123969NB-I00. The authors also acknowledge financial support through Grants No.\ PID2024-156352NB-I00 and No.\ PID2024-159024NB-C21, funded by MCIU/AEI/10.13039/501100011033/FEDER, UE and from Grant No. GR24022 funded by the Junta de Extremadura (Spain) and by European Regional Development Fund (ERDF) “A way of making Europe”. We have run our simulations in the computing facilities of the Instituto de Computaci\'{o}n Cient\'{\i}fica Avanzada de Extremadura (ICCAEx).

\begin{appendix}
\numberwithin{equation}{section}

\section{Discretized equation in terms of local fluctuations $u_i$}\label{appendix_ui}

The discrete-time evolution equation for $h_i$ reads
\begin{eqnarray}
h_{i}^{n+1} &=& h_{i}^n + \nu\Delta t\sum_{j\ne i} \left(h_{j}^n - h_{i}^n\right) \nonumber\\
    & & + \frac{\lambda\Delta t}{2}\sum_{j\ne i} \left(h_j^n - h_i^n\right)^2
    + \sqrt{2D\Delta t}\hspace{1mm}\eta_i^n \, .
\end{eqnarray}
Introducing fluctuations around the spatial mean,
\begin{equation}
u_i^n = h_i^n - \bar h^n , \qquad
\bar h^n = \frac{1}{N}\sum_j h_j^n ,
\end{equation}
the linear term reduces to
\begin{equation}
\sum_{j\neq i}(h_j^n-h_i^n)
= -N u_i^n .
\end{equation}

The quadratic term can be rewritten as
\begin{align}
\sum_{j\neq i}(h_j^n-h_i^n)^2
&= \sum_{j\neq i}(u_j^n-u_i^n)^2 \nonumber\\
&= \sum_{j\neq i}
\left[(u_j^n)^2 + (u_i^n)^2 - 2u_j^n u_i^n\right] \nonumber\\
&= \sum_j (u_j^n)^2 + (N-2)(u_i^n)^2
- 2u_i^n \sum_{j\neq i} u_j^n .
\end{align}
Since the fluctuations satisfy $\sum_j u_j^n = 0$, it follows that
$\sum_{j\neq i} u_j^n = -u_i^n$, and therefore
\begin{equation}
\sum_{j\neq i}(h_j^n-h_i^n)^2
= \sum_j (u_j^n)^2 + N (u_i^n)^2 .
\end{equation}
Defining the squared interface width
\begin{equation}
w^2 = \frac{1}{N}\sum_j (u_j^n)^2 ,
\end{equation}
the update rule becomes
\begin{equation}
h_i^{n+1}
= h_i^n
- \nu \Delta t\, N u_i^n
+ \frac{\lambda \Delta t\, N}{2}
\left[ w^2 + (u_i^n)^2 \right]
+ \sqrt{2D \Delta t}\,\eta_i^n .
\end{equation}
The evolution of the mean height follows from spatial averaging,
\begin{equation}
\label{eq:evolution_mean}
\bar h^{\,n+1}
= \bar h^{\,n}
+ \lambda \Delta t\, N w^2
+ \sqrt{2D \Delta t}\,\xi^{\,n},
\quad
\xi^{\,n}=\frac{1}{N}\sum_i\eta_i^n .
\end{equation}
Subtracting the mean, we finally obtain the time-evolution equation for the fluctuations,
\begin{equation}
\begin{aligned}
u_i^{n+1}
&= u_i^n - \nu \Delta t\, N u_i^n
+ \frac{\lambda \Delta t\, N}{2}
\left[ (u_i^n)^2 - w^2 \right] \\
&\quad + \sqrt{2D \Delta t}
\left( \eta_i^n - \xi^{\,n} \right).
\end{aligned}
\end{equation}

\section{Two-time correlations in terms of the local fluctuations $u_i$}
\label{appendix_aging_formula}

Starting from Eq.~\eqref{eq:aging}
\begin{equation}
\label{eq:aging_app1}
C_t(t,t_0)=\langle h_i(t)h_i(t_0)\rangle-\langle h_i(t)\rangle \langle h_i(t_0)\rangle,
\end{equation}
and using $h_i(t)=u_i(t)+\bar h(t)$, we obtain
\begin{align}
\langle h_i(t)h_i(t_0)\rangle
&=
\Big\langle \big[u_i(t)+\bar h(t)\big]\big[u_i(t_0)+\bar h(t_0)\big]\Big\rangle
\nonumber\\
&=
\langle u_i(t)u_i(t_0)\rangle
+\langle \bar h(t) u_i(t_0)\rangle
+\langle \bar h(t_0) u_i(t)\rangle
+\langle\bar h(t)\bar h(t_0)\rangle.
\label{eq:aging_ui1}
\end{align}
Likewise,
\begin{align}
\langle h_i(t)\rangle \langle h_i(t_0)\rangle
&=
\big[\langle u_i(t)+\bar h(t)\rangle\big]
\big[\langle u_i(t_0)+\bar h(t_0)\rangle\big]
\nonumber\\
&=
\big[\langle u_i(t)\rangle+\langle\bar h(t)\rangle\big]
\big[\langle u_i(t_0)\rangle+\langle\bar h(t_0)\rangle\big]
\nonumber\\
&=
\langle u_i(t)\rangle\langle u_i(t_0)\rangle
+\langle\bar h(t)\rangle\langle u_i(t_0)\rangle
+\langle\bar h(t_0)\rangle\langle u_i(t)\rangle
+\langle\bar h(t)\rangle\langle\bar h(t_0)\rangle.
\label{eq:aging_ui2}
\end{align}
The terms involving $\langle u_i\rangle$ vanish because, for each realization,
$\sum_i u_i(t)=0$, and all sites are statistically equivalent on the complete graph. Indeed,
\begin{equation}
0=\left\langle \sum_{i=1}^N u_i(t)\right\rangle
=\sum_{i=1}^N \langle u_i(t)\rangle
=N\langle u_i(t)\rangle,
\end{equation}
hence $\langle u_i(t)\rangle=\langle u_i(t_0)\rangle=0$.

The cross-terms vanish for the same reason. Indeed, for each realization
\begin{equation}
\sum_{i=1}^N u_i(t)=0,
\end{equation}
hence,
\begin{equation}
0=
\left\langle \bar h(t_0)\sum_{i=1}^N u_i(t)\right\rangle
=
\sum_{i=1}^N
\left\langle \bar h(t_0)u_i(t)\right\rangle .
\end{equation}
Since all sites are statistically equivalent on the complete graph, all terms in the last sum
are equal, and
\begin{equation}
0=
N\left\langle \bar h(t_0)u_i(t)\right\rangle .
\end{equation}
Thus, $\left\langle \bar h(t_0)u_i(t)\right\rangle=0$. Analogously, one also obtains
$\left\langle \bar h(t)u_i(t_0)\right\rangle=0$. Therefore, using these identities in Eqs.~\eqref{eq:aging_ui1} and \eqref{eq:aging_ui2}, one obtains
\begin{equation}
\langle h_i(t)h_i(t_0)\rangle=
\langle u_i(t)u_i(t_0)\rangle
+\langle\bar h(t)\bar h(t_0)\rangle,
\end{equation}
and
\begin{equation}
\langle h_i(t)\rangle \langle h_i(t_0)\rangle=
\langle\bar h(t)\rangle\langle\bar h(t_0)\rangle.
\end{equation}
Subtracting both expressions, one gets
\begin{align}
C_t(t,t_0)&=
\langle u_i(t)u_i(t_0)\rangle+\left[\langle\bar h(t)\bar h(t_0)\rangle-\langle\bar h(t)\rangle\langle\bar h(t_0)\rangle\right]
\nonumber\\
&=\langle u_i(t)u_i(t_0)\rangle+\mathrm{Cov}[\bar h(t),\bar h(t_0)].
\end{align}
The term $\mathrm{Cov}[\bar h(t),\bar h(t_0)]$ represents the contribution of the spatially uniform mode, namely the fluctuations of the global displacement of the interface. On the complete graph, $\bar h(t)$ is a spatial average over $N$ statistically equivalent nodes. Such global averages are self-averaging: the typical fluctuations of the average are smaller than those of each individual contribution, scaling as $N^{-1/2}$. Therefore, the corresponding covariance scales as $N^{-1}$.

This contribution is thus subleading in the large-$N$ limit. By contrast, the correlations of $u_i(t)$ describe local fluctuations around the instantaneous mean interface position. These local fluctuations are not spatial averages and are therefore not suppressed by the same self-averaging mechanism. Therefore, in the thermodynamic limit,
\begin{equation}
C_t(t,t_0)=\langle u_i(t)u_i(t_0)\rangle ,
\end{equation}
which is Eq.~\eqref{eq:aging_3}. In practice, within our numerical accuracy, we do not observe appreciable finite-size effects associated with the zero-mode contribution, indicating that it is already negligible for the system sizes that we have considered.

\section{Roughness of the EW equation on a complete graph}\label{appendix_EW_w}

Starting from the EW equation defined on a complete graph,
\begin{equation}
\frac{d h_i}{dt}
= \nu \sum_{j \neq i} (h_j - h_i) + \eta_i(t),
\label{eq:ew_raw}
\end{equation}
and writing the differential equation for the fluctuations $u_i$, following the steps outlined in Appendix~\ref{appendix_ui}, one arrives at
\begin{equation}
\label{eq:ui_dif}
\frac{d u_i}{dt}
= -\nu N u_i + \zeta_i(t),
\end{equation}
where the effective noise is given by
\begin{equation}
\zeta_i(t) = \eta_i(t) - \xi(t),
\qquad
\xi(t) = \frac{1}{N} \sum_j \eta_j(t).
\end{equation}

We now introduce a set of independent Wiener processes $\{W_k(t)\}$ satisfying
\begin{equation}
\langle dW_k \rangle = 0,
\qquad
\langle dW_k \, dW_l \rangle = \delta_{kl} \, dt,
\end{equation}
and represent the Gaussian white noise in differential form as
\begin{equation}
\eta_i(t) \, dt = \sqrt{2D} \, dW_i(t).
\end{equation}
It then follows that
\begin{equation}
\xi(t) \, dt = \frac{\sqrt{2D}}{N} \sum_k dW_k(t),
\end{equation}
and therefore
\begin{equation}
\zeta_i(t) \, dt =\sqrt{2D}\left(dW_i(t) - \frac{1}{N} \sum_k dW_k(t)\right).
\end{equation}

The corresponding It\^o stochastic differential equation then reads
\begin{equation}
d u_i= -\nu N u_i \, dt+ \sqrt{2D}\left(dW_i - \frac{1}{N} \sum_{k=1}^N dW_k\right).
\end{equation}
Since the noise terms are Gaussian, this equation may alternatively be written in terms of an effective single Wiener process as
\begin{equation}
d u_i= -\nu N u_i \, dt+ \sqrt{2D\left(1 - \frac{1}{N}\right)} \, dW_i.
\end{equation}

Applying It\^o’s lemma \cite{Gardiner2009} to compute the differential of $u_i^2$, one obtains
\begin{equation}
\label{eq:integracion_Ito}
\begin{aligned}
    d(u_i^2)=&\left[-2 \nu N u_i^2+ 2D \left(1 - \frac{1}{N} \right)\right] dt+\\
    & + \left[2 u_i \sqrt{2D \left(1 - \frac{1}{N} \right)}\right] dW_i.
\end{aligned}
\end{equation}
Taking averages over the noise realizations and using the fact that the stochastic term has zero mean yields
\begin{equation}
d \langle u_i^2 \rangle=\left[-2 \nu N \langle u_i^2 \rangle+ 2D \left(1 - \frac{1}{N} \right)\right] dt.
\end{equation}
By taking spatial averaging and recalling that the global roughness is defined as $w^2(t) = \langle \overline{u_i^2} \rangle$, we finally obtain the evolution equation
\begin{equation}
\frac{d}{dt} w^2(t) =-2 \nu N w^2(t)+ 2D \left(1 - \frac{1}{N} \right),
\end{equation}
whose solution, subject to the initial condition $w^2(0)=0$, is
\begin{equation}
\label{eq:rugosidad_EW_ape}
w^2(t)=\frac{D (N-1)}{\nu N^2}\left( 1 - e^{-2 \nu N t} \right).
\end{equation}

\section{Time power spectrum of the EW equation on a complete graph}\label{appendix_EW_ps}

Taking the spatial average of Eq.~\eqref{eq:ew_raw}, one obtains
\begin{equation}
\label{eq:ew_average}
\frac{d \bar{h}}{dt}
= \frac{1}{N} \sum_i \eta_i(t)
= \xi(t) .
\end{equation}
This equation for $\bar{h}(t)$ implies that the evolution of the average front position is Brownian. Since
\begin{equation}
\label{eq:ruido_ap_1}
\langle \eta_i(t)\,\eta_j(t') \rangle
= 2D\,\delta_{ij}\,\delta(t - t')\,,
\end{equation}
it follows that
\begin{equation}
\label{eq:ruido_ap_2}
\langle \xi(t)\,\xi(t') \rangle
= \frac{2D}{N}\,\delta(t - t')\,.
\end{equation}
By taking the time-Fourier transform of Eq.~\eqref{eq:ew_average}, one obtains
\begin{equation}
\label{eq:ew_fourier}
i\omega\tilde{\bar{h}}=\tilde{\xi}(\omega),
\end{equation}
where tildes are employed to denote transforms. Since 
\begin{equation}
\left\langle |\tilde{\xi}(\omega)|^2 \right\rangle=\int_{-\infty}^{\infty}dt\int_{-\infty}^{\infty}dt' e^{-i\omega(t-t')}\left\langle \xi(t)\xi(t')\right\rangle=\frac{2D}{N}\int_{-\infty}^{\infty}dt=\frac{4\pi D}{N}\delta(0), 
\end{equation}
where we have used the formal identity $\int_{-\infty}^{\infty}dt=2\pi\delta(0)$, one obtains
\begin{equation}
\label{eq:ew_ps}
S(\omega)=\langle|\tilde{\bar{h}}(\omega)|^2\rangle=\frac{4\pi\delta(0) D}{N\omega^2}\sim\omega^{-2}.
\end{equation}
For a finite observation window of duration $T$, this formally divergent factor is regularized as $\delta(0)=T/(2\pi)$. Thus, the EW contribution to the time power spectrum of the spatially averaged height decays as $\omega^{-2}$. For comparison with the simulations, however, one has to use the normalization of the discrete Fourier transform employed in the numerical analysis. In our case, the time series contains $M$ measurements separated by a time interval $\Delta t_{\mathrm{meas}}$, and the discrete frequencies are
\begin{equation}
\omega_q=\frac{2\pi q}{M}.
\end{equation}
With our DFT normalization, the corresponding prediction is
\begin{equation}
\label{eq:ew_ps_num}
S(\omega_q)
\simeq
\frac{4D\Delta t_{\mathrm{meas}}}{NM}\omega_q^{-2}.
\end{equation}

\section{Aging for the EW equation on a complete graph}\label{appendix_aging_EW}

From Eq.~\eqref{eq:ui_dif}, and using the integrating factor
$\mu(t)=e^{\nu N t}$, it is straightforward to show that the solution for $u_i(t)$ reads
\begin{equation}
u_i(t)
= e^{-\nu N t} u_i(0)
+ \int_0^t e^{-\nu N (t-s)} \zeta_i(s)\, ds .
\end{equation}
Moreover, $u_i(t)$ can be expressed in terms of its value at an earlier time $t_0$ as
\begin{equation}
u_i(t)
= e^{-\nu N (t-t_0)} u_i(t_0)
+ \int_{t_0}^t e^{-\nu N (t-s)} \zeta_i(s)\, ds .
\end{equation}
It follows that
\begin{equation}
\begin{aligned}
\langle u_i(t) u_i(t_0)\rangle
&= e^{-\nu N (t-t_0)} \langle u_i^2(t_0)\rangle \\
&\quad + \int_{t_0}^t e^{-\nu N (t-s)}
\left\langle u_i(t_0)\zeta_i(s)\right\rangle\, ds .
\end{aligned}
\end{equation}
Since $u_i(t_0)$ is independent of the noise $\zeta_i(s)$ for $s>t_0$
(because $u_i(t_0)$ depends only on the initial condition and on earlier noise realizations), the second term vanishes. Therefore,
\begin{equation}
\begin{aligned}
\overline{C_t(t,t_0)} &= \langle \overline{u_i(t) u_i(t_0)}\rangle = e^{-\nu N (t-t_0)} \langle \overline{u_i^2(t_0)}\rangle\\
&= w^2(t_0)\, e^{-\nu N (t-t_0)} .
\end{aligned}
\end{equation}

\section{Stability of the ST scheme}\label{appendix_estabilidad}

In this Appendix, we analyze why the ST integration scheme remains numerically stable and avoids overflows, despite its formal instability. Inspecting Eq.~\eqref{eq:integracion_ui}, two key aspects can be identified. First, the explicit Euler integration of the linear term reads
\begin{equation}
    u_i^{n+1} = (1 - \nu N \Delta t)\, u_i^n ,
\end{equation}
which is stable provided that $|1 - \nu N \Delta t| < 1$, or equivalently if $\Delta t < 2/(\nu N)$, with maximal stability achieved for $\Delta t \sim 1/(\nu N)$. This condition is always satisfied in our simulations, since $\Delta t$ is chosen to be much smaller. However, in the limit of very large system sizes, this constraint would eventually lead to numerical instabilities.

On the other hand, the nonlinear term is of the form $N (u_i^2-w^2)$. Since $u_i = w \chi_i$ and the roughness saturates as $w^2 \sim 1/N$, this contribution can be rewritten as being proportional to $\chi_i^2-1$. If the fluctuations $\chi_i$ are Gaussian, the largest contribution scales as $\chi_{\mathrm{max}}^2 \sim 2 \log N$, which can still be controlled by choosing a sufficiently small time step, typically $\Delta t \sim 1/N$.

Finally, we comment on why numerical runaways may occur for small system sizes $N$ even when $\Delta t$ is well below the linear stability bound. To isolate the mechanism, we neglect the stochastic forcing and focus on a large positive fluctuation $u_i$, for which the roughness correction is negligible ($u_i^2 \gg w^2$). The explicit Euler update then reduces to
\begin{equation}
u_i^{n+1}=u_i^n+N\Delta t\left(-\nu u_i^n+\frac{\lambda}{2}(u_i^n)^2\right).
\end{equation}
At the deterministic level, the drift changes sign at
\begin{equation}
u_c=\frac{2\nu}{\lambda},
\end{equation}
i.e.\ for $u_i>u_c$ the quadratic term dominates and the fluctuation is driven to grow, which can rapidly lead to numerical overflow. Since typical stationary fluctuations scale as $u\sim w\sim N^{-1/2}$, threshold-crossing events become more likely at small $N$. For example, for $\nu=1$ and $\lambda=20$, one has $u_c=0.1$, comparable to the typical fluctuation amplitude $w\simeq 0.1$ at $N=100$. In the full stochastic dynamics, rare noise-induced excursions can still drive individual sites above $u_c$ even at larger $N$. However, as $N$ increases these threshold-crossing events become progressively less likely; when they do occur, they can trigger a runaway.

\section{Average front position for the KPZ equation in the stationary state}\label{appendix_KPZ_h}

The KPZ equation can be rewritten, following the steps presented in Appendix~\ref{appendix_ui}, as
\begin{equation}
\frac{d h_i}{dt}
= -\nu N u_i
+ \frac{\lambda N}{2}\left[ w^2 + (u_i)^2 \right]
+ \eta_i(t).
\end{equation}
Taking the spatial average of this equation, one obtains
\begin{equation}
\label{eq:kpz_average}
\frac{d \bar{h}}{dt}
= \lambda N w^2 + \xi(t).
\end{equation}
Taking averages over noise realizations, one obtains:
\begin{equation}
\frac{d \langle \bar{h} \rangle}{dt}
= \lambda N w^2(t) .
\end{equation}
In the stationary state, $w^2(t)\approx D/(\nu N)$, yielding
\begin{equation}
\frac{d \langle \bar{h} \rangle}{dt}
\approx\frac{\lambda D}{\nu},
\end{equation}
hence, for the corresponding times
\begin{equation}
\langle \bar{h} \rangle \approx \frac{\lambda D}{\nu} t .
\end{equation}

\end{appendix}







\end{document}